\def\zem{$z_{\rm em}$}
\def\zabs{$z_{\rm abs}$}

\def\sii{S~{\sc ii}}

\def\hei{He~{\sc i}}
\def\heii{He~{\sc ii}}
\def\heiii{He~{\sc iii}}

\def\hi{H~{\sc i}}
\def\hii{H~{\sc ii}}

\def\cii{C~{\sc ii}}

\def\feii{Fe~{\sc ii}}

\def\siii{Si~{\sc ii}}
\def\ari{Ar~{\sc i}}

\def\alii{Al~{\sc ii}}

\def\aliii{Al~{\sc iii}}
\def\znii{Zn~{\sc ii}}

\def\oi{O~{\sc i}}

\documentclass[useAMS,usenatbib,epsfig]{mn2e}

\usepackage{rotating}
\usepackage{float}
\usepackage{natbib}
\usepackage{verbatim}
\usepackage{hyperref,graphicx}
\usepackage{pdflscape}

\title[On the deficiency of Argon in DLA systems]{The ESO UVES Advanced Data Products Quasar Sample -\\ IV. On the deficiency of Argon in DLA systems}
\author[T. Zafar et al.] {Tayyaba Zafar$^{1, 2}$\thanks{e-mail:tzafar@eso.org}, Giovanni Vladilo$^3$, C\'{e}line P\'{e}roux$^2$, Paolo Molaro$^3$, 
\newauthor
Miriam Centuri\'on$^3$, Valentina D'Odorico$^3$, Kumail Abbas$^4$, \& Attila Popping$^5$\\
$^1$ European Southern Observatory, Karl-Schwarzschild-Strasse 2, 85748, Garching, Germany. \\
$^2$ Aix Marseille Universit\'e, CNRS, LAM (Laboratoire d'Astrophysique de Marseille) UMR 7326, 13388, Marseille, France.  \\
$^3$ INAF - Osservatorio Astronomico di Trieste, via G.B. Tiepolo 11, Trieste, Italy.\\
$^4$ Center of Excellence in Solid State Physics, University of the Punjab, Lahore-54590, Pakistan. \\
$^5$ International Centre for Radio Astronomy Research (ICRAR), The University of Western Australia, 35 Stirling Hwy, Crawley,\\ WA 6009, Australia.\\
}

\begin{document}


\pagerange{\pageref{firstpage}--\pageref{lastpage}} \pubyear{2013}

\maketitle

\label{firstpage}

\begin{abstract}
In this work, we study argon abundances in the interstellar medium of high-redshift galaxies
($2.0 \leq$ \zabs\ $\leq 4.2$) detected as Damped Ly$\alpha$ absorbers (DLA) in the spectra of background quasars. We use high-resolution quasar spectra obtained from the ESO-UVES advanced data products (EUADP) database. We present 3 new measurements and 5 upper limits of \ari . We further compiled DLAs/sub-DLA data from the literature with measurements available of argon and $\alpha$-capture
elements (S or Si), making up a total of 37 systems, i.e. the largest DLA argon sample investigated so far. We confirm that argon is generally deficient in DLAs, with a mean value [Ar/$\alpha$] $\simeq -0.4\pm 0.06$\,dex (standard error of the mean). The [Ar/$\alpha$] ratios show a weak, positive trend with increasing \hi\ column density and increasing absorption redshift, and a weak, negative trend with dust-free metallicity, [S/H]. Detailed analysis of the abundance ratios indicates that \ari\ ionisation, rather than dust depletion or nucleosynthetic evolution, is responsible for the argon deficiency. Altogether, the observational evidence is consistent with a scenario of argon ionisation dominated by quasar metagalactic radiation modulated by local \hi\ self-shielding inside the DLA host galaxies. 
Our measurements and limits of argon abundances suggest that the cosmic reionisation of \heii\ is completed above $z \sim 3$, but more measurements at \zabs\ $>$ 3.5 are required to probe the final stages of this process of cosmic reionisation.
\end{abstract}

\section{Introduction}

High-resolution quasar spectra show a large number of absorption lines 
produced in the diffuse gas intercepted by the line of sight,
from redshift $z=0$ up to the quasar's redshift.
Most quasar absorption lines originate in the intergalactic medium (IGM)
and contain information on the physical properties 
of a large volume of the Universe over a large fraction of look-back time. 
A small fraction of quasar absorption systems, 
those with the highest values of neutral hydrogen column density, 
N(\hi) $\geq 10^{20.3}$ atoms cm$^{-2}$,
originate in the diffuse gas of intervening galaxies. 
These absorbers, called Damped Ly$\alpha$ absorbers (DLAs) \citep{wolfe86,wolfe05},
contain information on the physical state and chemical composition 
of the interstellar medium (ISM) in high-redshift galaxies.
By studying DLAs we can cast light on the astrophysical processes that take place
in their host galaxies and, in particular, on the interaction between the metagalactic radiation 
propagating through the IGM and the interstellar gas at high redshift. 
In this work, we focus our attention on a particular type of absorption lines
in DLAs, the absorption of neutral argon, which offers a way 
to probe this interaction. 

Owing to the high value of its first ionisation potential ($15.76$\,eV), argon is expected
to be mostly neutral in interstellar \hi\ regions, 
which are opaque to photons with energies just above 
the \hi\ ionisation threshold ($13.6$\,eV).
Nevertheless, the ionisation fraction of argon 
is very sensitive to high energy ionising photons 
that are able to leak through the neutral gas.
This sensitivity is due to the high ratio of 
 photoionisation to recombination rates, 
 typically one order of magnitude higher 
for \ari\ than for \hi\ \citep{sofia98}. 
The fraction of \ari\ is expected to vary according to the energy
distribution of the radiation field. 
When the radiation field is hard, \ari\ is predicted to be deficient
 relative to other ``low ions'' typical of \hi\ regions. 
On the other hand, normal (i.e. solar) abundances are expected when the ionising spectrum is soft. 
In our own Galaxy, neutral argon is found to be deficient in low-density interstellar regions
\citep{sofia98,jenkins00,lehner03,jenkins13}.
However, owing to the saturation of \ari\ lines, Galactic studies can only probe regions
with relatively low \hi\ column density, typically associated to warm interstellar gas. 
On the other hand, thanks to the low metallicity of DLA galaxies \citep[e.g.][]{wolfe05},
\ari\ lines can be unsaturated in DLAs
in spite of the high column densities of these absorbers,
N(\hi) $> 10^{20.3}$ atoms cm$^{-2}$ \citep{wolfe86}.
Therefore, \ari\ lines offer a way 
to probe the ionisation state of the neutral gas in high-redshift galaxies detected
in absorption. 
%
%
In principle, 
by measuring the ionisation state of the gas at different redshifts, we can track the
evolution of the comoving density of cosmic sources and variations of the transparency of the IGM 
resulting from the processes of cosmic reionisation. 

According to our current understanding of the reionisation history of the Universe
\citep{fan06, mcquinn09,jarosik11,haardt12}, 
hydrogen was reionised by the ultraviolet radiation
of the first stars and quasars, starting at $z \geq 10$;
with the increasing overlap of IGM \hii\ regions, hydrogen reionisation
was completed around $z \geq 6$ ,
leaving the IGM completely transparent to photons with $h\nu > 13.6$\,eV.
%
During the early stages of \hi\ reionisation, the ionising radiation field was
sufficiently hard to ionise \hei\ ($IP=24.6$\,eV),
but not \heii\ ($IP=54.4$\,eV). 
At a later stage, when the density of hard UV-emitting quasars became sufficiently high,
also \heii\ was reionised. This process possibly started around $z\sim6$ \citep{bolton12},
leading to the complete overlap of IGM \heiii\ regions 
around $z \simeq 3$ \citep{furlanetto08,mcquinn09}.
With the completion of \heii\ reionisation the IGM became completely transparent to 
photons with $h\nu > 54.4$\,eV. 

Probing the processes of cosmic reionisation 
requires a variety of experimental tools.
Observations of \ari\ in DLAs provide a rare possibility to
probe the effects of internal and external radiation fields acting inside high-redshift galaxies.
Since the stellar radiation originated inside the galaxy is softer than that of the quasar background, 
one can in principle discriminate the ionisation sources 
by taking advantage of the sensitivity of \ari\ to high energy photons. 
This type of study helps to understand the last stages of cosmic reionisation
that take place in a redshift interval where DLAs can be easily observed.

Neutral argon can be measured from the resonant transitions at 1048\,\AA\ and 1066\,\AA.
In DLAs these lines are observable from ground in the visible and near ultraviolet 
when the absorption redshift is \zabs\ $\geq$ 2. Argon lines are difficult to detect because
they can be weak in the DLAs of lower metallicity and because they fall in the Ly$\alpha$ forest of the spectrum. 
The first measurement of \ari\ abundance in a DLA system,
namely the absorber at \zabs=3.39 towards Q\,0000-2620, did not show evidence
for a deficiency of argon \citep{molaro01}. 
The first systematic study, based on a sample of 10 measurements and 5 limits of \ari, 
indicated that the absorber at \zabs=3.39 towards Q\,0000-2620 is exceptional, since argon is deficient
in most DLAs (\citealt{vladilo03}; hereafter Paper\,I). 
An evolutionary trend between argon deficiency and redshift was found
in the redshift interval covering the epoch of \heii\ reionisation.
However, the limited size of the sample made difficult to derive firm conclusions on this and other
trends of argon abundances.

In the present work we reassess the current status of argon abundances
in DLAs after the accumulation of many years of observations obtained 
with the European Southern Observatory (ESO) Ultraviolet-Visual Echelle Spectrograph (UVES).
This spectrograph, optimized for the near ultraviolet \citep{dekker00}
and fed by the Unit 2 of the Very Large Telescope (VLT), is
an ideal instrument for this type of research. 
To enlarge the sample of argon measurements, we use 
the Advanced Data Products (EUADP) sample \citep{zafar13,zafar13b},
together with an updated compilation of literature data. 
The new sample and the analysis of the data are presented in \S 2 and notes on new discoveries are provided in \S 3. In \S 4 we present the results, showing that
argon is generally deficient in DLAs
\S 5 we discuss the nature of argon deficiency and its possible causes and in \S 6 we discuss the nature of the ionising sources and the evidence of evolution. The conclusions of our work are provided in \S 7.

\section{Data sample}
\subsection{The EUADP data}
This work is based on high-resolution ($\lambda/\Delta \lambda$$\simeq$$5 \times 10^4$) quasar spectra obtained from the ESO-UVES instrument and collected in the EUADP sample. Details on building and processing of the EUADP sample are presented in \citet{zafar13}. This EUADP sample of 250 quasar spectra has been used to study DLAs/sub-DLAs \citep{zafar13b}. Indeed, 197 DLAs and sub-DLAs have been reported in these quasar spectra, thus providing the ideal parent sample for the study of rare elements in quasar absorbers. In particular, the EUADP sample allows for the observations of \ari\ at redshifts 1.86 $<$ $z$ $<$ 4.74. In this work, we search for such \ari\ lines in 140 quasar absorbers. \ari\ is not detected in more than one-third of the cases. This is due to $i)$ the lack of wavelength coverage in the relevant region, $ii)$ the presence of a Lyman Limit System, or $iii)$ low signal-to-noise ratio in the region of interest. Moreover, 9 measurements, 1 upper and 1 lower limit of \ari\ have been already reported in the literature for the EUADP DLAs/sub-DLAs (see Table \ref{argon}). Of the remaining systems, more than half were rejected due to strong blending of \ari\ with the lines from the Ly$\alpha$ forest. We find 3 new detections of \ari\ never reported before. In addition to these measurements, we report 5 new upper limits of \ari\ in this EUADP DLAs/sub-DLAs sample.

\subsection{Fitting method}
The ionic column densities (\ari, \sii, \siii, etc) were derived using the $\chi^2$-minimization routine \texttt{FITLYMAN} within the \texttt{MIDAS} environment \citep{fontana95}. Laboratory wavelengths and oscillator strengths were taken from \citet{morton03}. The global fit returns the best fit parameters for the central wavelength, column density and Doppler turbulent broadening, as well as 1$\sigma$ errors on each of these quantities.

Given that the two lines of \ari\ transitions are located in the Ly$\alpha$ forest, great care was taken to assess the contamination of the lines by interlopers. This possible blending would tend to overestimate the column density of \ari. A comparison with radial velocity profiles of low ionisation species (\sii, \siii, and/or \feii) has been made to uncover the transitions which are free from contamination. These low ionisation species were chosen as reference absorption lines because they lie outside the Ly$\alpha$ forest and are thus free from possible blending. The agreement with the radial velocity profiles of the \ari\ lines is generally satisfactory. Upper limits were derived in the cases with evidences for contamination of \ari\ transitions by Ly$\alpha$ forest interlopers. This is done by fixing the $b$ and $z$ parameters of \ari\ lines to those of the other detected low ionisation lines and fitting for column density. The details of each individual system are given in \S \ref{indv:obj}.

\subsection{Literature data}
In addition to the measurements presented here, we made an investigation in the literature for quasar absorbers with \ari\ abundance determination. We find 17 measurements, 11 upper and 1 lower limit of \ari . Incidentally, some of the objects are part of the EUADP DLA/sub-DLA sample. The combined sample makes a total of 37 DLAs/sub-DLAs (20 measurements together with 16 upper and 1 lower limit). The targets, the basic data for the DLAs/sub-DLAs investigated, and \ari\ total column densities are listed in Table \ref{argon} (see also Table \ref{tab:sample}).

\addtocounter{table}{+1}
\begin{table}
\caption{Summary of the number of systems with detected \ari\ in the two samples.}
\centering
\label{tab:sample}
\begin{tabular}{l c c c}
\hline\hline
Sample & Measurements & Upper limits & Lower limits \\
\hline
EUADP & 3 & 5 & $\cdots$ \\
Literature & 17 & 11 & 1 \\
\hline
\end{tabular}
\end{table}

\section{Notes on individual objects}
\label{indv:obj}
In the following, we briefly describe each system for which we report here for the first time the \ari\ abundance. Additional information on the UVES spectra and the determination of \hi\ column densities for these absorbers can be found in \citet{zafar13,zafar13b}.

\begin{figure}
  \centering
{\includegraphics[width=\columnwidth,clip=]{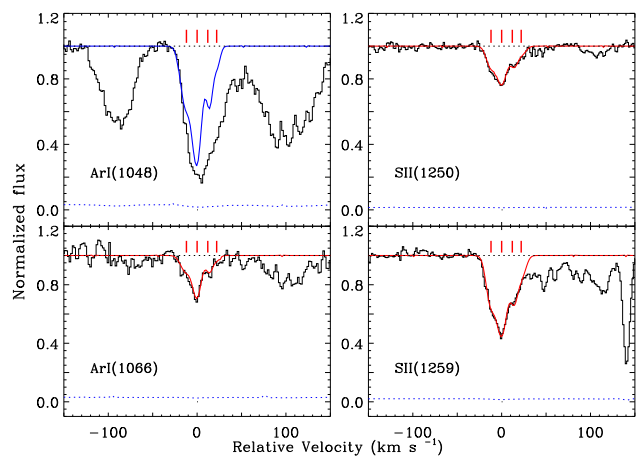}}
     \caption{Voigt-profile fits (red overlay) of \ari\ and \sii\ for the DLA at \zabs=3.2655 (zero velocity) in J\,1113-1533. Normalised quasar spectrum and error spectrum are shown in black and blue dotted lines, respectively. The blue overlay represents the profile of the transition which is not fit and obtained from the fit of the other transitions in the system. The locations of the voigt profile components} are indicated by the red tick marks.
     \label{fig:1113}
  \end{figure}

\underline{J\,1113-1533, \zabs=3.2655:} this DLA has an \hi\ column density of log N(\hi) $=21.23\pm0.05$. The four-component structure of this system is derived from the fit to the \sii\ $\lambda\lambda$ 1250, 1259 absorption lines. The argon abundance is derived from unsaturated, non-blended \ari\ $\lambda$ 1066 line (see Fig. \ref{fig:1113}). The redshift of the components, Doppler parameters and column densities are listed in Table \ref{tab:1113}. The \siii\ lines are either saturated or blended. We derive a lower limit from \siii\ $\lambda\lambda$ 1304, 1526 transitions: log N(\siii)$>15.08$ by using the redshift and $b$-value of the components obtained from the \sii\ transitions.

\begin{table}
\caption{Results of voigt profile decomposition of the \zabs=3.2655 DLA towards J\,1113-1533.}
\centering
\label{tab:1113}
\begin{tabular}{l c c c c}
\hline\hline
Comp. & \zabs\ & $b$ & Ion & log N \\
	 & & km s$^{-1}$ & & cm$^{-2}$ \\
\hline
1 & 3.2653 & $5.6\pm1.5$ & \sii\ & $13.95\pm0.02$ \\
    &              &                         & \ari\ & $12.93\pm0.03$ \\
2 & 3.2655 & $6.9\pm1.7$ & \sii\ & $14.35\pm0.02$ \\
    &              &                         & \ari\ & $13.30\pm0.02$ \\
3 & 3.2657 & $3.9\pm0.9$ & \sii\ & $13.86\pm0.02$ \\
    &              &                         & \ari\ & $12.78\pm0.04$ \\
4 & 3.2658 & $1.9\pm0.8$ & \sii\ & $13.46\pm0.02$ \\
    &              &                         & \ari\ & $12.12\pm0.04$ \\
\hline
\end{tabular}
\end{table}

\begin{figure}
  \centering
{\includegraphics[width=\columnwidth,clip=]{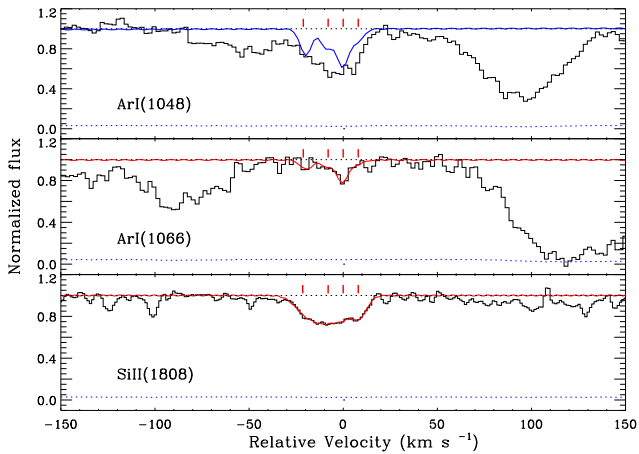}}
     \caption{Voigt-profile fits (red overlay) of \ari\ and \siii\ for the DLA at \zabs=2.9669 (zero velocity) in J\,1356-1101.}
\label{fig:1356}
 \end{figure}

\underline{J\,1356-1101, \zabs=2.9669:} this DLA has an \hi\ column density of log N(\hi) $=20.80\pm0.10$. The four-component structure of the system is derived from the fit to the \siii\ $\lambda$ 1808 absorption line. The argon abundance is derived from unsaturated, non-blended \ari\ $\lambda$ 1066 line (see Fig. \ref{fig:1356}). The fit parameters are described in Table \ref{tab:1356}. Note that \citet{noterdaeme08} derived log N(\siii)$=14.96\pm0.06$, consistent with our results within the errors.

\begin{table}
\caption{Results of voigt profile decomposition of the \zabs=2.9669 DLA towards J\,1356-1101.}
\centering
\label{tab:1356}
\begin{tabular}{l c c c c}
\hline\hline
Comp. & \zabs\ & $b$ & Ion & log N \\
	 & & km s$^{-1}$ & & cm$^{-2}$ \\
\hline
1 & 2.9666 & $6.2\pm1.5$ & \siii\ & $14.33\pm0.04$ \\
    &              &                         & \ari\ & $12.86\pm0.04$ \\
2 & 2.9668 & $7.4\pm1.7$ & \siii\ & $14.59\pm0.03$ \\
    &              &                         & \ari\ & $12.47\pm0.03$ \\
3 & 2.9669 & $2.1\pm0.9$ & \siii\ & $14.02\pm0.04$ \\
    &              &                         & \ari\ & $13.15\pm0.03$ \\
4 & 2.9670 & $3.2\pm1.1$ & \siii\ & $14.28\pm0.03$ \\
    &              &                         & \ari\ & $12.21\pm0.03$ \\
\hline
\end{tabular}
\end{table}

\begin{figure}
  \centering
{\includegraphics[width=\columnwidth,clip=]{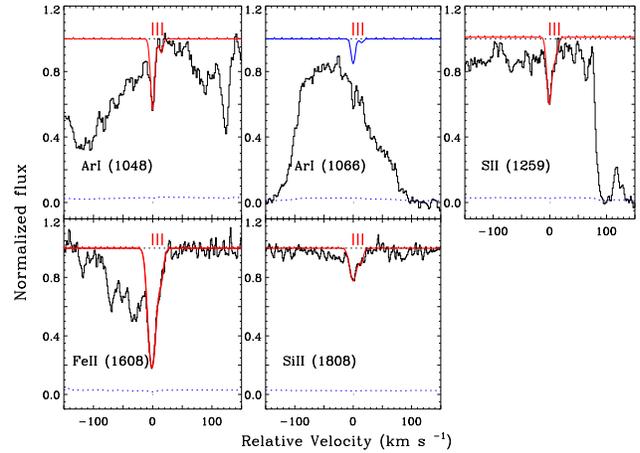}}
     \caption{Voigt-profile fits (red overlay) of \ari, \sii, \feii, and \siii\ (see labels) for the DLA at \zabs=3.0572 (zero velocity) in B\,2332-094.}
\label{fig:2332}
  \end{figure}

\underline{B\,2332-094, \zabs=3.0572:} this sub-DLA has an \hi\ column density of log N(\hi) $=20.27\pm0.07$. The metal profile is spread over several components. However, unsaturated, non-blended \ari\ $\lambda$ 1048 line is seen in only three of the strongest components (see Fig. \ref{fig:2332}). 
This three-component structure is used to fit the \sii\ $\lambda$ 1259, \feii\ $\lambda$ 1608, and \siii\ $\lambda$ 1808 absorption lines. The parameters for the fit are summarised in Table \ref{tab:2332}. Note that \citet{petitjean08,prochaska03} derived log N(\sii)$=14.34\pm0.18$, log N(\feii)$=14.34\pm0.03$, and log N(\siii)$=14.86\pm0.07$ by fitting all the components in this system. Logically, our determinations of the total column densities are slightly lower than the one reported by other authors.

\begin{table}
\caption{Results of voigt profile decomposition of the \zabs=3.0572 DLA towards B\,2332-094.}
\centering
\label{tab:2332}
\begin{tabular}{l c c c c}
\hline\hline
Comp. & \zabs\ & $b$ & Ion & log N \\
	 & & km s$^{-1}$ & & cm$^{-2}$ \\
\hline
1 & 3.0572 & $5.8\pm1.5$ & \sii\ & $14.03\pm0.02$ \\
    &              &                         & \siii\ & $14.54\pm0.04$ \\
    &              &                         & \feii\ & $14.01\pm0.02$ \\
    &              &                         & \ari\ & $12.90\pm0.03$ \\
2 & 3.0573 & $1.8\pm0.5$ & \sii\ & $13.36\pm0.03$ \\
    &              &                         & \siii\ & $12.87\pm0.04$ \\
    &              &                         & \feii\ & $12.42\pm0.02$ \\
    &              &                         & \ari\ & $11.53\pm0.03$ \\
3 & 3.0574 & $2.3\pm0.7$ & \sii\ & $12.70\pm0.02$ \\
    &              &                         & \siii\ & $13.93\pm0.04$ \\
    &              &                         & \feii\ & $12.97\pm0.02$ \\
    &              &                         & \ari\ & $11.92\pm0.03$ \\\hline
\end{tabular}
\end{table}

\section{Argon abundances in DLA systems} \label{AbundancePatterns}

In this section we discuss the elemental abundances of argon obtained
from the measurements presented in Table \ref{argon}. 
We follow the common convention expressing the number abundance ratios 
of two elements X and Y
according to the relation 
$[\mathrm{X}/\mathrm{Y}] = 
\log_{10} (\mathrm{X}/\mathrm{Y})_\mathrm{gas} - \log_{10} (\mathrm{X}/\mathrm{Y})_\mathrm{ref}$,
where $(\mathrm{X}/\mathrm{Y})_\mathrm{gas}$ is the ratio of the column densities measured in the gas phase
and $(\mathrm{X}/\mathrm{Y})_\mathrm{ref}$ is a reference abundance based on the solar composition.
Here we adopt the set of solar photospheric abundances of \citet{asplund09}.

Unfortunately, it is difficult to obtain accurate solar abundances for ``noble gas'' elements
such as argon. The value that we adopt here,
$\xi(\mathrm{Ar}) \equiv \log_{10} (\mathrm{Ar}/\mathrm{H}) +12 = 6.40\pm0.13$, 
is an average of results obtained from a variety of techniques \citep{asplund09}.
In Paper\,I, we adopted the solar abundances of \citet{grevesse98}
for most elements and the argon solar value proposed by \citep{sofia98}, $\xi(\mathrm{Ar}) = 6.52$, 
which is marginally consistent with the value proposed by \citet{asplund09}. Taking into account these differences, 
the [Ar/S] and [Ar/Si] ratios that we calculate now are higher by 
+0.04\,dex and +0.07\,dex, respectively, with respect to those calculated in Paper\,I. The uncertainty of the reference value affects the level of argon abundance that we derive, but not the correlation trends that we present below.

  \begin{figure*}
  \centering
\includegraphics[width=\columnwidth,clip=]{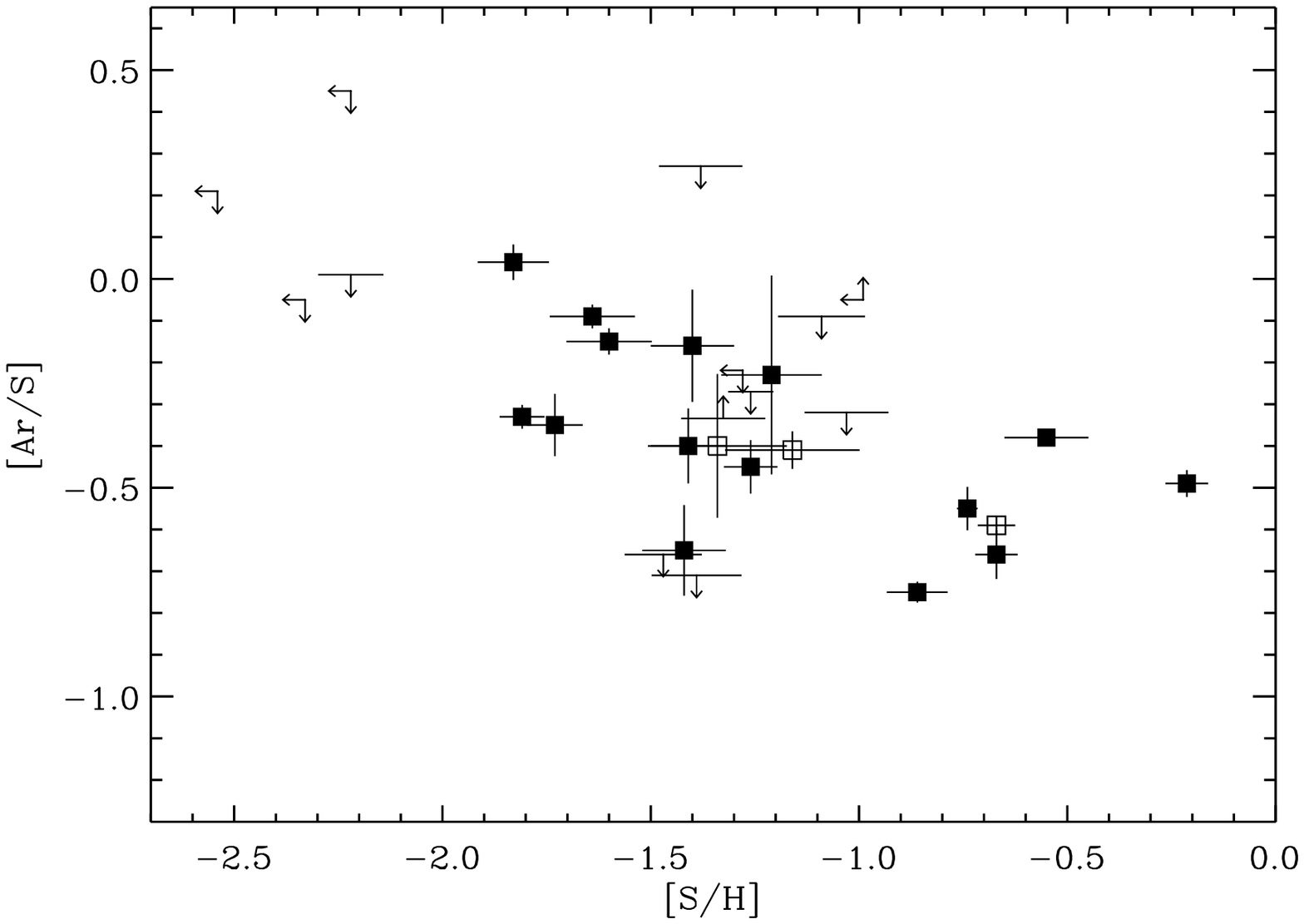}
\includegraphics[width=\columnwidth,clip=]{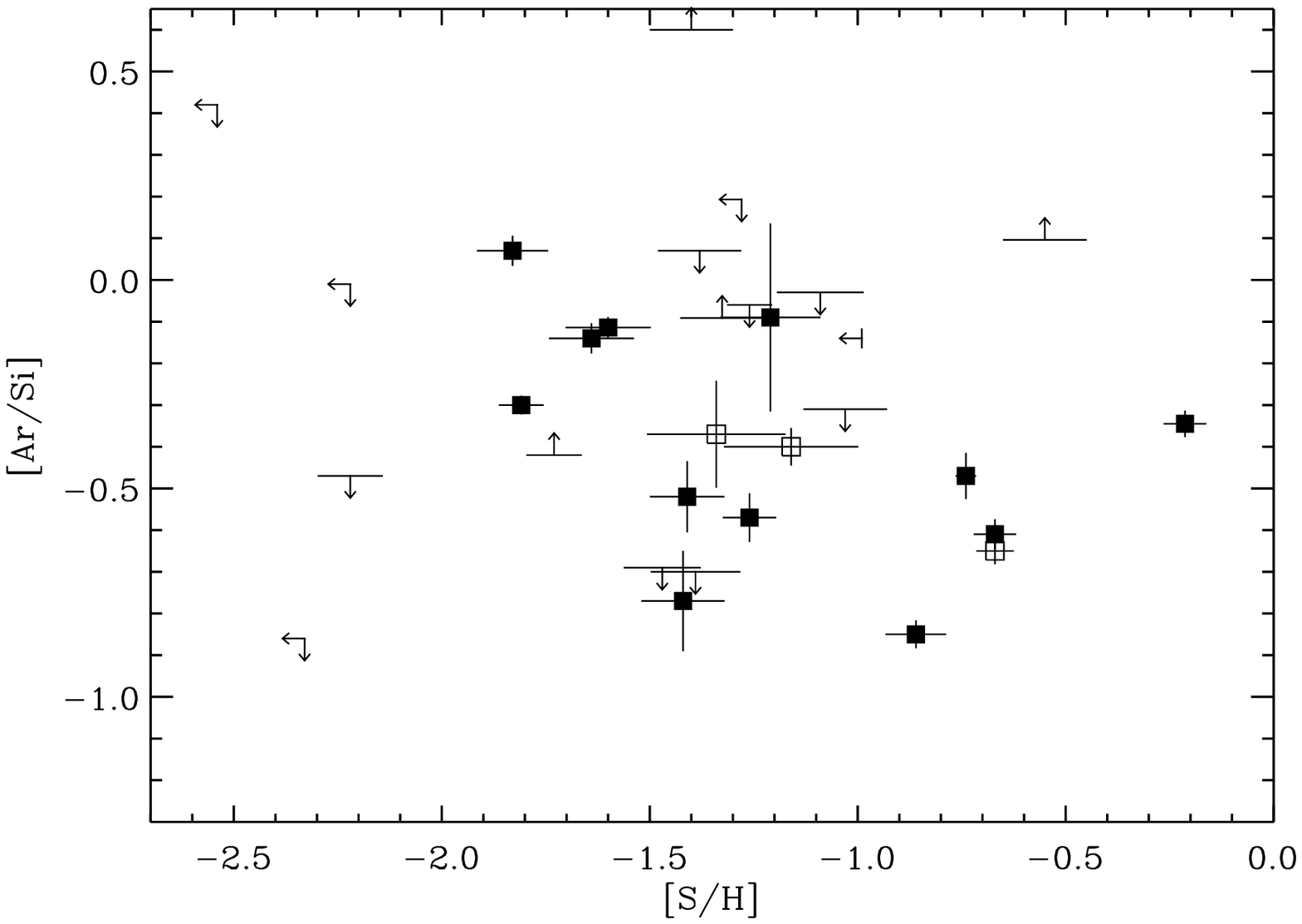}
     \caption{The [Ar/S] and [Ar/Si] versus S-based metallicities in DLA systems. Solid squares and arrows correspond to measurements and limits, respectively. The open squares represent proximate DLAs. A small chemical evolution could be seen in both [Ar/S] and [Ar/Si].}
         \label{arsmet}
  \end{figure*}

\subsection{Argon deficiency in DLA systems} \label{sectArgonDeficiency} 

In our sample, the absolute abundance of argon
(i.e., the abundance relative to hydrogen)
spans the range $-2.4\,\mathrm{dex} \leq $ [Ar/H] $\leq -0.4\,\mathrm{dex}$, with a mean value
$\langle [\mathrm{Ar}/\mathrm{H}] \rangle = -1.57 \pm 0.08$\,dex (standard error of the mean; 20 measurements).
Broadly speaking, this result is in line with the fact that DLA systems are metal poor.
However, the mean value [Ar/H] is lower than the mean metallicity of DLA
systems, [M/H] $\simeq -1.1$\,dex \citep{wolfe05}, indicating
that argon is underabundant with respect to other metals.
To evidentiate this underabundance, we 
use relative abundances, i.e., the abundances of argon relative to other products of stellar nucleosynthesis.
By using relative abundances, we can offset differences in the level of metallicity attained
by individual DLA galaxies, and we can 
focus our attention on effects other than metallicity, such as dust and ionisation effects. 
In practice, we use sulphur and silicon as reference elements to measure the relative abundances. 
The existence of a relatively large number of \sii\ and \siii\ 
column density measurements in DLA systems is one reason for choosing these two elements.
In addition, silicon and sulphur have nucleosynthetic history similar to that of argon
and negligible (or low) dust depletion (see \S \ref{OriginArgonDeficiency}).

 From the analysis of  our sample of measurements, rather than limits, we obtain the following results. 
 The relative abundance with respect to sulphur spans the range
$-0.75\,\mathrm{dex} \leq$ [Ar/S] $\leq +0.04\,\mathrm{dex}$, 
with mean value $\langle [\mathrm{Ar/S}] \rangle = -0.39\pm0.05$\,dex (standard error of the mean; 18 systems). 
Using silicon as a reference, we find values 
 $-0.85\,\mathrm{dex} \leq$ [Ar/Si] $\leq +0.07\,\mathrm{dex}$, with a mean value
$\langle [\mathrm{Ar/Si}] \rangle = -0.39\pm0.06$\,dex (17 systems).

The upper limits are consistent with these values
and in most cases are stringent enough to confirm the general deficiency of argon. 
Therefore, the results obtained from the updated 
 \ari\ sample confirm the general deficiency of argon in DLA systems
found in Paper I.

The standard deviation of the argon abundance ratios
is $0.21$\,dex and $0.25$\,dex for [Ar/S] and [Ar/Si], respectively. 
The spread of these abundance ratios is somewhat higher than that
of other relative abundances in DLA systems. 
For instance, we find a $1\sigma$ scatter of 0.14\,dex and 0.17\,dex 
for [Zn/S] and [Si/Fe], respectively,
for the current sample of 
DLA systems with column densities
of \sii, \znii, \siii\ and \feii\ obtained from high-resolution spectra.

\subsubsection{Proximate DLA systems}

A few absorption systems of our sample are proximate DLAs, 
i.e. their relative velocity separation from the background quasar is 
$\Delta v < 3000$\,km s$^{-1}$ \citep[e.g.,][]{moller98,ellison10,zafar11}.
Specifically, these are
the absorber at $z_{abs}=2.811$ towards B\,0528-2505, at $z_{abs}=3.174$ towards J\,1337$+$3152, and 
that at $z_{abs}=3.167$ towards J\,1604+3951. 
These systems will be treated separately in the subsequent analysis
since they may not be representative of the typical population of DLA systems
\citep{ellison10}. 
With the exclusion of these three cases,
the mean values and standard errors of the mean are
$\langle [\mathrm{Ar/S}] \rangle = -0.37\pm0.06$\,dex (15 systems)
and $\langle [\mathrm{Ar/Si}] \rangle= -0.38\pm0.07$\,dex (14 systems), confirming that argon is significantly deficient in classic DLA systems.

\subsection{Trends of argon abundances} \label{ObservedTrends}

The relatively large sample of the present study allows us to 
reassess the presence of trends between the relative abundances of argon
and other quantities that we can measure in DLAs.
In particular, we review the evidence for trends  
with metallicity, hydrogen column density and absorption redshift.
In Tables \ref{correlations_full} and \ref{correlations} we provide the results 
of a correlation analysis performed 
for the complete sample and the sample without proximate DLAs, respectively. 
In these Tables we give the Pearson's correlation coefficient, $\rho_\mathrm{PE}$
and the Spearman's rank correlation coefficient, $\rho_\mathrm{SP}$,
obtained from the datasets of [Ar/S] and [Ar/Si] measurements. 
We also give the  probability of null hypothesis, $p_\mathrm{PE}$ and $p_\mathrm{SP}$
obtained for the Pearson's and Spearman's tests, respectively.
The $p$-values roughly indicate the probability of an uncorrelated population producing datasets that have a Pearson 
or Spearman correlation at least as extreme as the one computed from the measured datasets. 

\subsubsection{Trend with metallicity}

The absorption lines of the \sii\ triplet at 1250\,\AA\ 
are usually detected in the quasar spectra where \ari\ is detected.
For this reason, and because sulphur is not affected by dust depletion (see \S \ref{DustDepletion}),
we use the absolute abundance of sulphur, [S/H], as a metallicity indicator in our study. 

In Fig. \ref{arsmet} we plot our measurements and limits of [Ar/S] and [Ar/Si] versus [S/H].
Visual inspection shows that argon becomes more deficient as the metallicity increases.
At the lowest values of metallicity of our sample, the relative abundances are
approximately solar, whereas at the highest metallicities they are 
deficient by at least $\sim -0.5$\,dex with respect to the solar value.  

Some evidence for an anti-correlation between [Ar/S] and metallicity is present in both samples
(Tables \ref{correlations_full} and \ref{correlations}),
with correlation coefficients $\rho_\mathrm{PE} \simeq -0.6$ 
and null hypothesis probabilities $p_\mathrm{PE} \simeq 1\%$.  
A similar trend is found for [Ar/Si], albeit with a weaker correlation,
($\rho_\mathrm{PE} \simeq -0.5$; $p_\mathrm{PE} \simeq 6\%$/$14\%$).
The Spearman tests yield  similar results. 
The existence of a trend between argon abundances and metallicity 
 in DLAs is reported here for the first time.

\subsubsection{Trend with hydrogen column density}

In Fig. \ref{arsnh} we plot our measurements and limits of [Ar/S] and [Ar/Si] versus log N(\hi).
Visual inspection shows that argon has a tendency to become more abundant as the \hi\ column density increases,
albeit with a large spread. 
At the lowest \hi\ column densities of our sample the relative abundances are deficient by more than $\sim -0.5$\,dex , whereas 
at the highest column densities they may attain approximately solar values. 
 
The statistical analysis of [Ar/S] versus log N(\hi) for the complete sample
indicate some evidence for a positive correlation
($\rho_\mathrm{PE} \simeq +0.6$; $p_\mathrm{PE} \simeq 2\%$). 
The evidence is stronger for the
sample without proximate DLAs ($\rho_\mathrm{PE} \simeq +0.7$; $p_\mathrm{PE}\simeq 0.5\%$).
A positive correlation is also present for [Ar/Si], but less robust:
we find $\rho_\mathrm{PE} \simeq +0.4$ ($p_\mathrm{PE}\simeq 15\%$) for the complete sample,
and $\rho_\mathrm{PE} \simeq +0.5$ ($p_\mathrm{PE} \simeq 6\%$) for the sample without proximate DLAs.
The Spearman tests yield similar results, with slightly larger null-hypothesis  probabilities 
for the  [Ar/Si] versus log N(\hi) trends. 

In the sample of Paper\,I there was no statistical evidence for 
the existence of a correlation with N(\hi), even though the
DLA with highest N(\hi) did show the highest argon abundance.

\subsubsection{Trend with absorption redshift}

In Fig. \ref{arsz} we plot our measurements and limits of [Ar/S] and [Ar/Si] versus \zabs.
Visual inspection shows that argon has a tendency to become more abundant as the redshift increases,
but with a large spread. The statistical analysis of [Ar/S] versus \zabs\
shows a very weak evidence for a positive correlation
($\rho_\mathrm{PE} \simeq +0.2$; $p_\mathrm{PE}\simeq 41\%$) for both samples
(with and without proximate DLAs). 
A similar, positive trend is found for the [Ar/Si] ratios, with a slightly higher correlation:
we find $\rho_\mathrm{PE} \simeq +0.35$ ($p_\mathrm{PE}\simeq 16\%$) for the complete sample
and $\rho_\mathrm{PE} \simeq +0.43$ ($p_\mathrm{PE}\simeq 12\%$) for the sample without proximate DLAs. 
The Spearman tests yield slightly lower correlation coefficients and larger probabilities of null-hypothesis.  
The weakness of the correlations  with absorption redshift 
contrasts with the high confidence level found
in Paper\,I for the same type of analysis. The larger size of the present sample reveals a more complex
situation, with a modest trend and a large scatter.

\begin{table}
\caption{Correlation analysis of argon abundances for the complete sample
of measurements in DLAs.}        
\label{correlations_full}    		
\begin{minipage}[t]{\columnwidth}
\centering                                  
\renewcommand{\footnoterule}{}  
\setlength{\tabcolsep}{5pt}
\begin{tabular}{c c | r c r c | r c r c r}
\hline\hline                       
$y$ & $x$ & $n$ & $\rho_\mathrm{PE}$ & $p_\mathrm{PE}$& $\rho_\mathrm{SP}$ &  $p_\mathrm{SP}$\\
\hline 
$\textrm{[Ar/S]}$ & $\textrm{[S/H]}$ & 18 & $-0.63$ & 0.5\% & $-0.65$ & 0.3\% \\
$\textrm{[Ar/Si]}$ & $\textrm{[S/H]}$ & 15 & $-0.49$ & 6.3\% & $-0.50$ & 5.6\%\\
$\textrm{[Ar/S]}$ & log N(\hi) & 18 & $+0.55$ & 1.8\% & $+0.55$ & 1.8\%\\
$\textrm{[Ar/Si]}$ & log N(\hi) & 17 & $+0.37$ & 14.7\% & $+0.32$ & 21.2\%\\
$\textrm{[Ar/S]}$ & \zabs\ & 18 & $+0.21$ & 40.9\% & +0.13 & 60.7\%\\
$\textrm{[Ar/Si]}$ & \zabs\ & 17 & $+0.35$ & 16.4\% & +0.21 & 41.4\%\\
\hline \hline
\end{tabular}
\\
\end{minipage}
\end{table}

\begin{table}
\caption{Correlation analysis of argon abundances
for the sample without proximate DLAs.}            
\label{correlations}    		
\begin{minipage}[t]{\columnwidth}
\centering                                  
\renewcommand{\footnoterule}{}  
\setlength{\tabcolsep}{5pt}
\begin{tabular}{c c | r c r c r}
\hline\hline                       
$y$ & $x$ & $n$ & $\rho_\mathrm{PE}$ & $p_\mathrm{PE}$& $\rho_\mathrm{SP}$ &  $p_\mathrm{SP}$ \\
\hline 
$\textrm{[Ar/S]}$ & $\textrm{[S/H]}$ & 15 & $-0.61$ & 1.7\% & $-0.64$ & 1.1\% \\
$\textrm{[Ar/Si]}$ & $\textrm{[S/H]}$ & 12 & $-0.45$ & 14.2\% & $-0.49$ & 10.6\%\\
$\textrm{[Ar/S]}$ & log N(\hi) & 15 & $+0.69$ & 0.5\% & $+0.69$ & 0.5\%\\
$\textrm{[Ar/Si]}$ & log N(\hi) & 14 & $+0.51$ & 6.3\% & $+0.45$ & 10.4\%\\
$\textrm{[Ar/S]}$ & \zabs\ & 15 & $+0.23$ & 40.1\% & +0.13 & 64.8\%\\
$\textrm{[Ar/Si]}$ & \zabs\ & 14 & $+0.43$ & 12.5\% & +0.22 & 45.4\%\\
\hline \hline
\end{tabular}
\\
\end{minipage}
\end{table}
 
\section{Nature of the argon deficiency} \label{OriginArgonDeficiency}

The abundance of argon, as any other elemental abundance
measured in DLAs, is influenced
by three types of effects at work in the diffuse gas of the host galaxy:
dust depletion, nucleosynthesis and ionisation processes. Disentangling the relative contribution of these effects is crucial for 
a correct interpretation of the measured abundances. 

%
%

\subsection{Dust Depletion} \label{DustDepletion}

\begin{figure*}
  \centering
\includegraphics[width=\columnwidth,clip=]{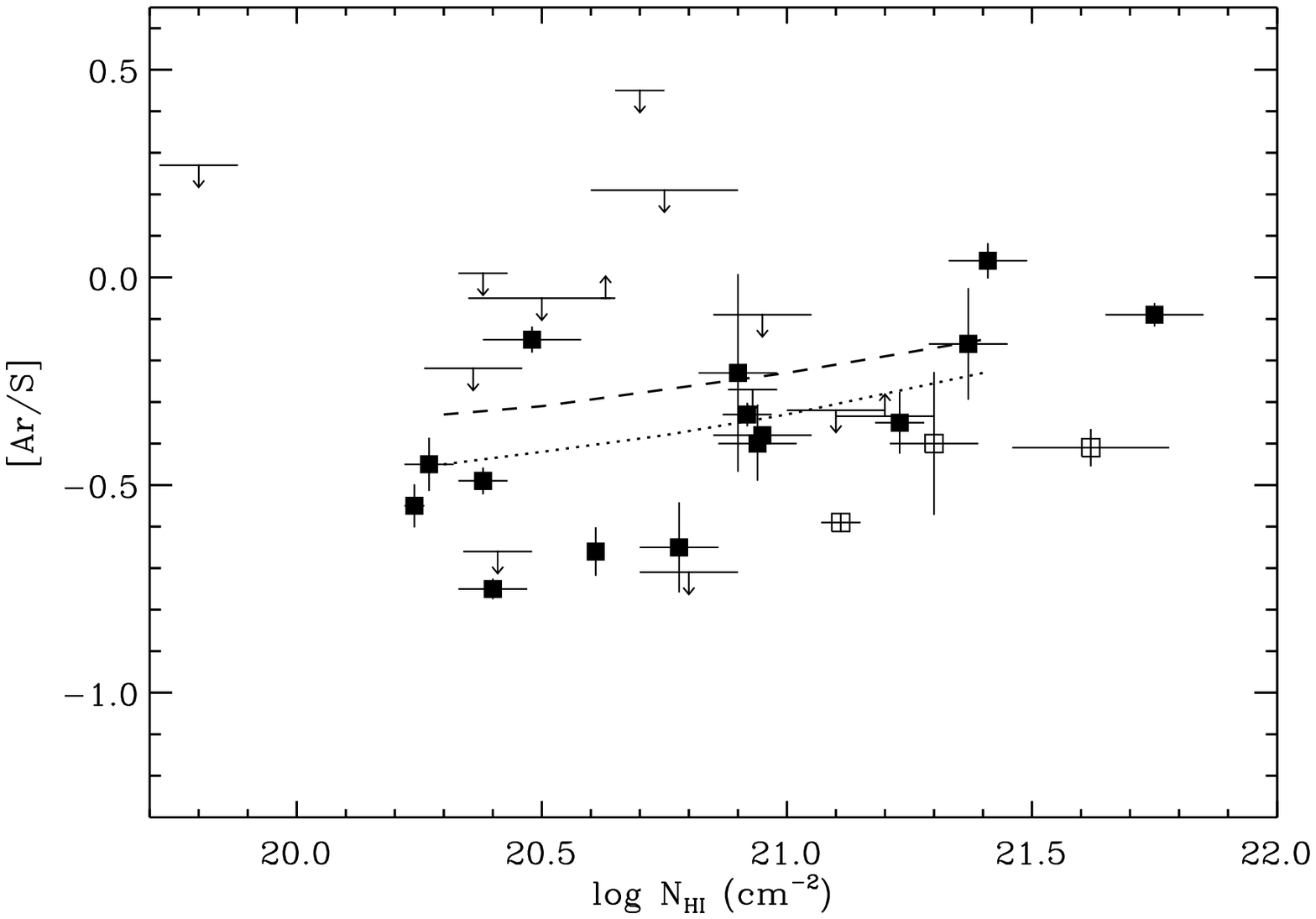}
\includegraphics[width=\columnwidth,clip=]{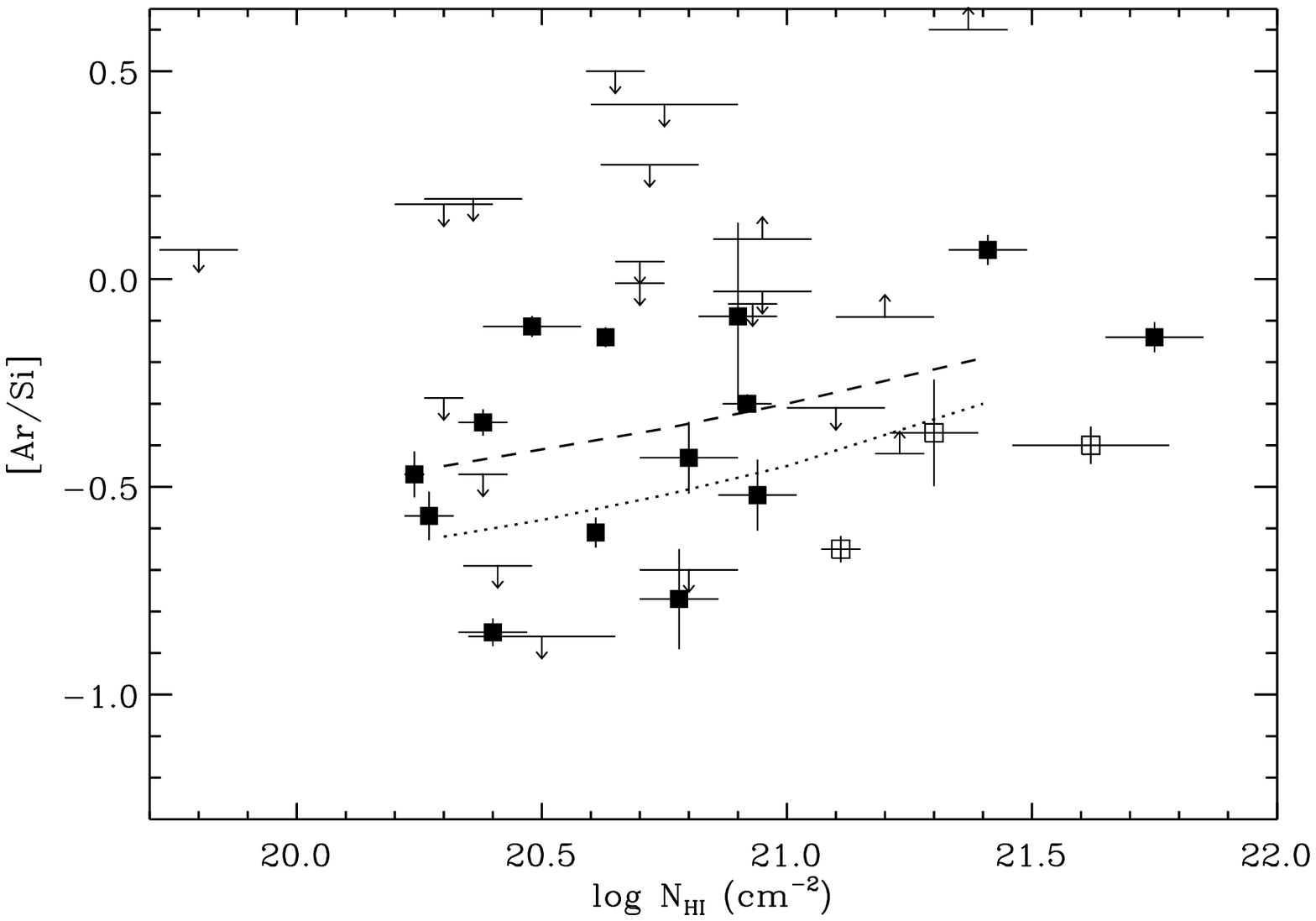}
     \caption{The [Ar/S] and [Ar/Si] ratio in DLA systems plotted versus \hi\ column density. Solid squares and arrows correspond to measurements and limits, respectively. The open squares illustrate proximate DLAs. Curves correspond to the predictions of idealized CLOUDY photoionisation models calculated for solar abundance ratios. Dashed curve: model predictions obtained for a layer of gas at $z=2.5$ $n_\mathrm{H}=0.1$ atoms cm$^{-3}$ embedded in the HM extragalactic radiation field \citep[e.g.][]{haardt12}; dotted curve: predictions obtained by doubling the intensity of the radiation field in the same model. See \S\ref{IonisationEffects}.}
  \label{arsnh}
 \end{figure*}

Dust depletion is due to the incorporation of part of the atoms of the ISM into solid phase (dust grains). Elements classified as volatiles and refractories according to condensation calculations \citep{lodders03} are characterized by low and high levels of depletion, respectively \citep{Savage96}. Dust depletion in DLAs is known to be moderate, similar to that found in low-density components of the Galactic ISM \citep{pettini94,pettini97,ledoux02,vladilo11}. For this reason, depletions of well-known volatile elements, such as zinc or sulphur, or even moderately refractory elements, such as silicon, have negligible or modest effects on DLA abundance measurements. However, depletions of refractory elements, such as iron, do affect the abundance measurements and will not be considered in our work, which is focussed on ionisation effects. Argon, as a ``noble gas'', has little tendency to take part to chemical reactions, including those that lead to the condensation of dust grains.
Detailed calculations of condensation temperatures indicate, indeed, that argon is an extremely volatile element \citep{lodders03}. In principle, argon atoms could form the stable hydride ArH$^+$ through the reaction Ar$^+$+H$_2 \rightarrow$ ArH$^+$+H \citep{duley80,sofia98}. However, this reaction requires molecular hydrogen, but the fraction of molecular hydrogen, $f(\mathrm{H_2}) \equiv 2 N(\mathrm{H_2})/[N($\hi$) + 2 N(\mathrm{H_2})]$, is generally small or negligible in DLAs \citep[see e.g.,][]{srianand12}. The absorber at $z_\mathrm{abs}=2.337$ towards B\,1232+0815 is the only system of our sample with $f(\mathrm{H_2}) \ga 0.1$ \citep{noterdaeme08}. This systems shows a modest argon depletion (see Table \ref{argon}), suggesting that the molecular route for argon depletion may require a high fraction of molecular gas. Therefore, we do not expect the ArH$^+$ production to play a role in the other molecular systems of our sample, which have at most $f(\mathrm{H_2}) \simeq 0.01$ (see Table \ref{argon}). 

To test whether argon is depleted or not from an experimental point of view, in Fig. \ref{znfe} we plot the ratio [Ar/S] versus [Zn/Fe]. The ratio [Zn/Fe] is a well known indicator of dust depletion in the ISM \citep[e.g.,][]{Savage96}
commonly employed in studies of DLA systems (e.g., \citealt{pettini94,pettini97,ledoux02,ellison09,vladilo11}; see however \citealt{rafelski12}). If argon is depleted into dust, we would expect a trend of decreasing [Ar/S] with increasing [Zn/Fe], assuming sulphur undepleted. Fig. \ref{znfe} shows instead a large scatter without a clear trend. A linear regression analysis of [Ar/S] versus [Zn/Fe] for the sample without proximate DLA systems yields a Pearson's correlation coefficient $\rho=-0.20$  ($p=63\%$; 8 systems). Therefore, even if the statistics are too small to draw firm conclusions, there is no evidence of trend. 

Based on the theoretical considerations discussed above, and given the lack of a trend with [Zn/Fe], in the rest of the discussion we assume that the dust depletion of argon is negligible.


\subsection{Nucleosynthesis \label{sectNucleosynthesis}}

The abundance ratios measured in DLAs tend to change
in the course of the evolution of the host galaxies
as a result of the enrichment of the diffuse gas by heavy elements
produced by stellar nucleosynthesis and ejected by supernova explosions. Studies of Galactic chemical evolution indicate that the relative abundances
of elements with a similar nucleosynthetic origin, such as $\alpha$-capture elements,
undergo a modest evolution as the global level of metallicity increases. 
Therefore, relative abundance changes due to ionisation effects for the $\alpha$-elements S, Si and Ar considered here are expected to be small.
Unfortunately, to our knowledge, there are no measurements of argon abundances in the samples of metal-poor stars
used to trace the effects of Galactic chemical evolution. Argon abundances can be measured in early-type stars \citep[e.g.,][]{Lanz08},
but these stars are short-lived and therefore not suitable for tracing the chemical evolution of the Galaxy. Observational studies of argon chemical evolution
are based on measurements of emission lines from \hii\ regions
rather than on stellar measurements. By comparing measurements obtained in extragalactic \hii\ regions 
of different metallicity,
\citet{henry99} find that abundance ratios Ne/O, S/O, and Ar/O 
appear to be universally constant and independent of metallicity. Similar conclusions are obtained from a study of O, Ar and S abundances
in metal-poor \hii\ regions \citet{izotov99}. 
From the analysis of the data published in Table 1 of that paper,
and adopting the solar reference abundances of \citet{asplund09}, we find
a mean value $\langle [\mathrm{Ar/S}] \rangle = +0.02\, (\pm 0.08)$\,dex,
without evidence of evolution of [Ar/S] with metallicity [S/H]. 
Based on these observational results,
we conclude that nucleosynthetic effects are unable to explain
the general under-abundance of [Ar/$\alpha$] that we observe in DLAs. 

 \begin{figure*}
  \centering
\includegraphics[width=\columnwidth,clip=]{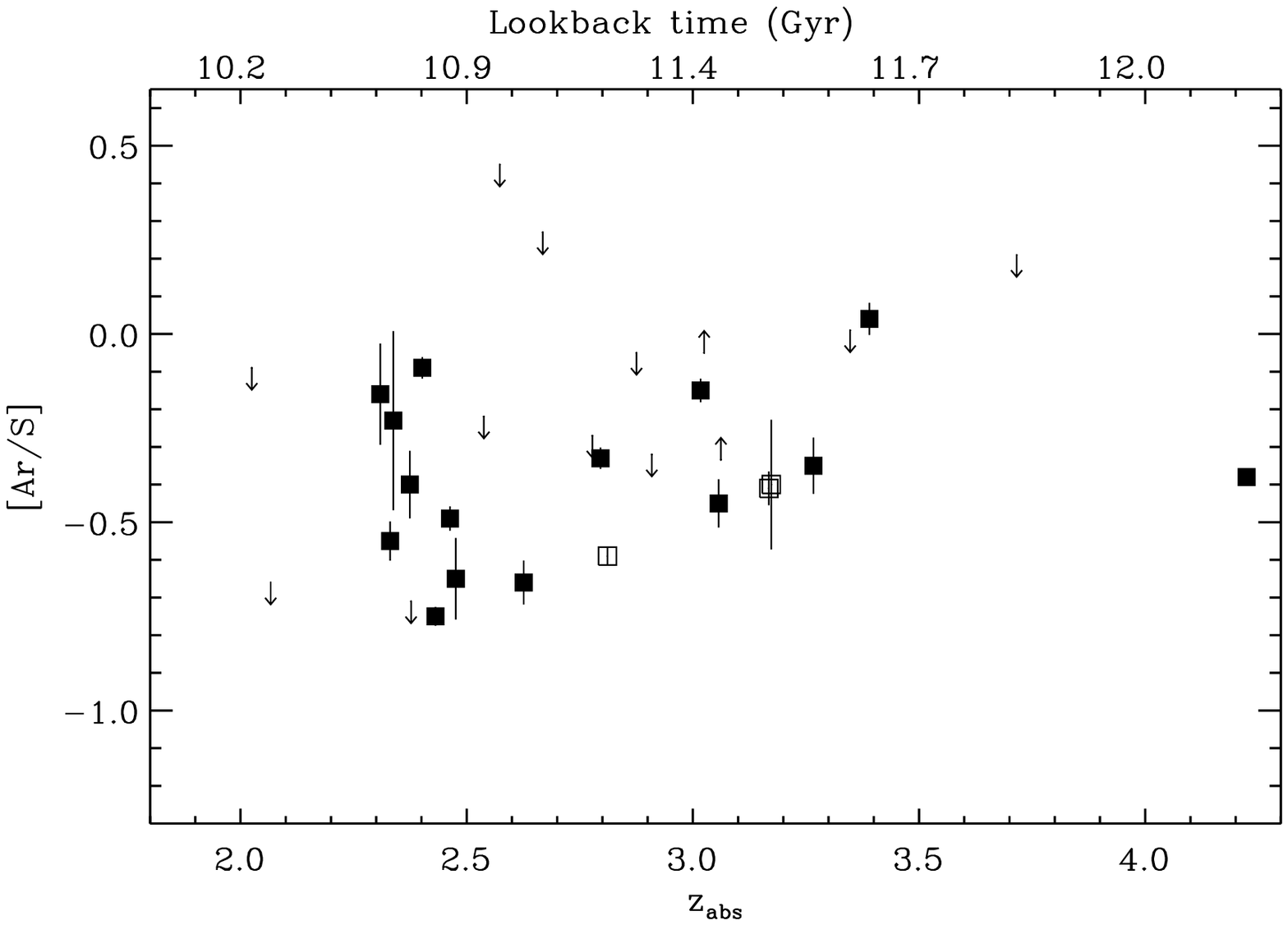}
\includegraphics[width=\columnwidth,clip=]{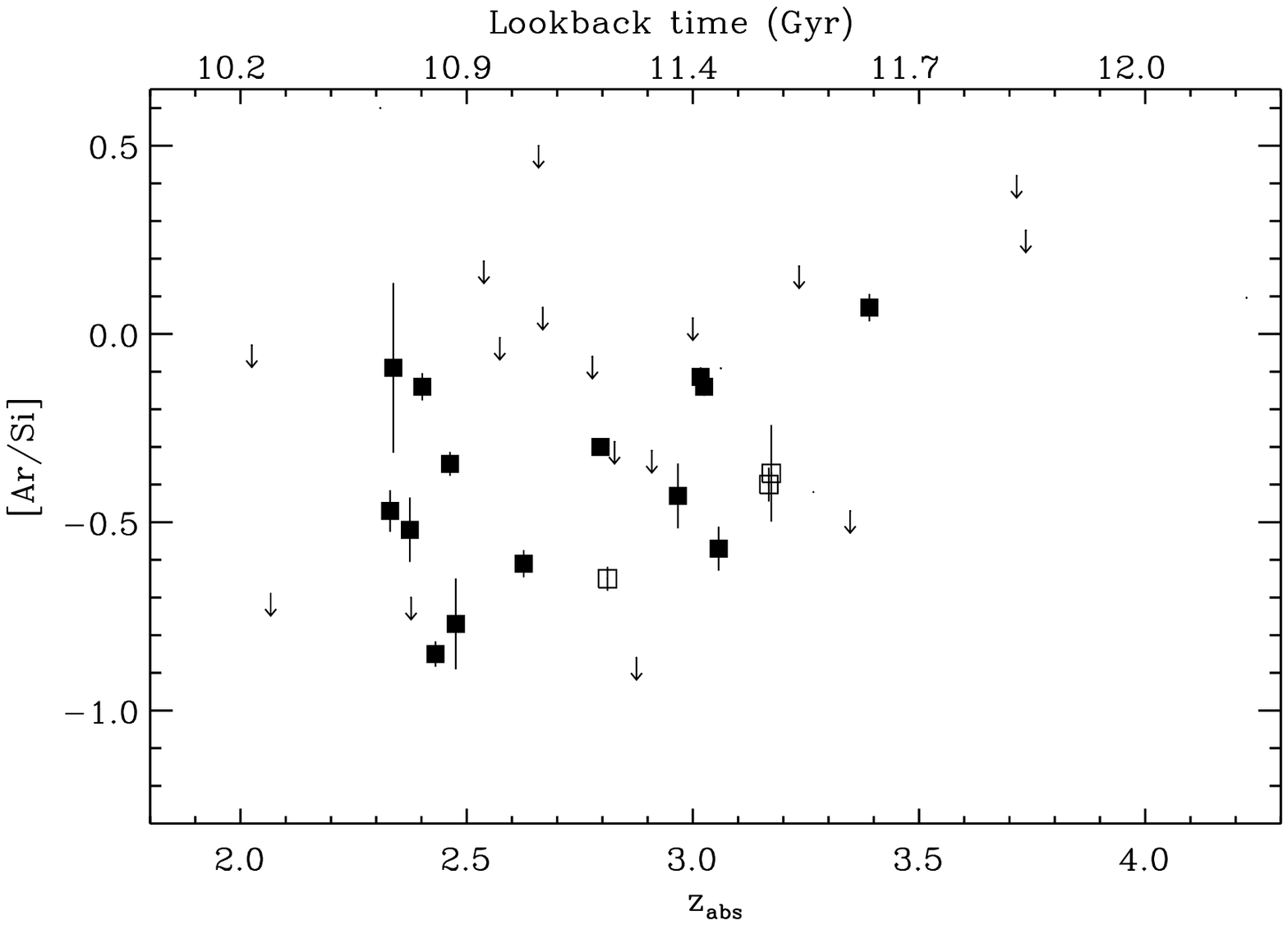}
     \caption{The [Ar/S] and [Ar/Si] ratios in DLA systems plotted versus redshift (bottom axis) and loockback time (top axis), for $H_0=70$ km s$^{-1}$ Mpc$^{-1}$, $\Omega_m=0.3$, and $\Omega_\Lambda=0.7$. The open squares correspond to proximate DLAs.}
         \label{arsz}
  \end{figure*}
%

 
\subsection{Ionisation effects} \label{IonisationEffects}

The argon abundance ratios are affected by photoionisation processes
induced by the radiation field that pervades the diffuse gas of DLA galaxies. 
The ionisation ratios measured in DLA systems can be modelled
with ionisation equilibrium codes, such as 
CLOUDY \citep{freland98}, assuming plane parallel geometry
with photons incident on one side of the cloud. We used CLOUDY (version 13.03; \citealt{ferland13}) to perform two sets of calculations: one with the Haardt \& Madau (HM) extragalactic background (Table HM05, see e.g. \citealt{haardt12}) at $z = 2.5$, and one with the radiation field of a star with $T_\mathrm{eff}=33,000$\,K \citep{kurucz91}, representative of stellar sources internal to the DLA host galaxy. In all calculations we adopted metal abundances 0.1 the solar reference values \citep{asplund09}. We included interstellar grains for heating and cooling processes with a dust-to-gas ratio 0.1 of the Galactic value, as well as cosmic ray heating typical of Galactic interstellar gas. We adopted a total hydrogen density $n_\mathrm{H}=0.1$ atoms cm$^{-3}$, a value representative of Galactic warm gas. The calculations were stopped at an assigned value of \hi\ column density. The highest values of \hi\ column density are difficult to model with CLOUDY since the deepest layers tend to become cooler than 1000\,K, the default value of the temperature threshold which is present in the code. To increase the possibility of tracking high density layers, we lowered the temperature threshold to 100\,K. To constrain the model, we required the predicted hydrogen ionisation to be $x \la 0.1$ and the \aliii/\alii\ ratio to match the observed data. To this end, we searched for measurements of \aliii\ and \alii\ column densities for the DLAs of our sample, checking that the absorption lines of both ions shared the same velocity profiles as the \ari\ lines. Unfortunately, only a few DLAs in our sample have \aliii\ and \alii\ measurements\footnote{This happens because the \alii\ line $\lambda$1676\AA\ is often saturated in DLAs. The few measurements of the \aliii/\alii\ ratio available only in our sample are indicated with the footnote ``c'' in Table \ref{argon}.}. Based on these data, we adopted log \aliii/\alii\ $\simeq -1$\,dex as a reference value. The fractions of \ari, \siii\ and \sii\ predicted by the ionisation models were used to calculate the [Ar/S] and [Ar/Si] ratios.

In the case of the extragalactic radiation field, the model was tuned by scaling the intensity of the HM field. Remarkably, the model is quite successful in matching the experimental data even taking the HM field at face value. At log N(\hi)=20.75\,dex the model yields 
$x=0.10$, log \aliii/\alii = $-1.1$ dex, [Ar/S] $=-0.27$\,dex and [Ar/Si]=$-0.36$\,dex.
The abundance of neutral argon tends to increase with increasing \hi\ column density.
The predictions of [Ar/S] and [Ar/Si] versus log N(\hi) are shown with a dashed line in Fig. \ref{arsnh}.  
One can see that the model yields values of [Ar/S] and [Ar/Si] in the range of
the observations and, in addition, predicts the existence of a trend
between argon deficiency and \hi\ column density consistent with the observed one 
(Fig. \ref{arsnh}).
The scatter of the experimental [Ar/S] and [Ar/Si] ratios at any value of N(\hi)
can be modelled, to some extent, invoking local variations of the extragalactic radiation field
and of the hydrogen density of the cloud.
As an example, we plot in Fig. \ref{arsnh} the results obtained 
by doubling the intensity of the HM radiation field (dotted line).
Similar results are obtained by decreasing $n_\mathrm{H}$ by a factor of two.
Variations of the cloud geometry, that cannot be accounted for by CLOUDY, 
may certainly introduce additional scatter. 

%

In the case of the stellar radiation field, the model was tuned by varying the ionisation parameter\footnote{
The ionisation parameter is the dimensionless ratio
$U=\Phi(\mathrm{H})/cn_\mathrm{H}$, where $\Phi(\mathrm{H})$ is the total surface flux
of ionising photons (cm$^{-2}$\,s$^{-1}$) and $n_\mathrm{H}$ the H particle density (cm$^{-3}$).}. However, with the stellar field CLOUDY is not able to finish the calculations because the gas temperature falls below the minimum allowed value before the total N(\hi) is attained. 
To bypass this problem, \citet{vladilo01} considered a two-region model composed of a 
mildly ionised interface and an inner neutral region 
that only hosts dominant ionisation species. With this type of model,
the CLOUDY calculations can be performed in the interface, thanks to its low \hi\ column density, 
and the ionisation ratios can be calculated by averaging the column densities over the two regions
(see \citealt{vladilo01} for more details). No significant argon underabundance  
was found with this model tuning the stellar field to match the measured \aliii/\alii\ ratios \citep{vladilo03}.
In the present work we have explored the predictions of the stellar field model as a function of depth in the cloud
in the range of \hi\ column densities for which the CLOUDY calculations are not interrupted. We find that the ionisation of Ar is negligible in the layers where the \aliii/\alii\ ratio attains values comparable to the observed ones.

In conclusion, ionisation processes offer a key for understanding the origin of the argon deficiency,
at variance with dust depletion and nucleosynthetic evolution effects.
The  general level of argon deficiency and the trend with \hi\ column density can be explained
by photoionisation models with radiation field dominated by extragalactic background. 
The fact that the stellar field is not able to produce a significant underabundance of argon must be related
to the paucity of stellar photons with energies of several tens of eV. In this energy range\footnote{
Photons in this energy range can penetrate the \hi\ layers because the \hi\ photoionisation cross section scales as $\nu^{-3}$
and therefore neutral layers tend to become transparent to ionising photons when $h\nu \gg 13.6$\,eV. },
where the spectrum of the extragalactic background does not decline,  
argon ionisation is particularly efficient because the ratio of photoionisation-to-recombination rates of \ari\ is high 
\citep[see, e.g.,][Fig. 3]{sofia98}.

\section{Sources of argon ionisation} \label{IonisingSources}

Based on the above discussion, 
we now assume that the measured \ari\ deficiency is due to ionisation  
and we investigate the nature of the ionising sources. 

\subsection{External sources}

The comparison between observational data and model predictions in Fig. \ref{arsnh} suggests that the metagalactic UV/X-ray background impinging on the DLA host galaxies is the most natural source of ionisation. The trend of Ar abundances with redshift (\S \ref{ObservedTrends}) provides an observational test for probing the hypothesis of external ionisation.

\subsubsection{Interpretation of the trend with redshift} \label{InterpretationRedshift}

In the redshift interval typical of most DLA observations ($2 \leq z_{\rm abs}\leq 3.5$) the hydrogen ionisation rate in the IGM has a weak tendency to decrease with redshift, albeit with large uncertainties \citep[see e.g. left panel of Fig. 8 in][]{haardt12}. The CLOUDY model with HM extragalactic radiation (\S\ref{IonisationEffects}) predicts the [Ar/S] and [Ar/Si] ratios to increase by 0.09\,dex and 0.13\,dex, respectively, when the redshift increases from $z = 2.0$ to $z = 3.5$ (case log N(\hi) $=20.75$\,dex). The measurements shown in Fig. \ref{arsz} and the correlation analysis in Tables \ref{correlations_full} and \ref{correlations} are consistent with a modest rise of argon abundances with redshift, in line with the model predictions. The large scatter observed at any given redshift can be naturally explained in terms of attenuation of the external field by \hi\ absorption inside the cloud.
This possibility is supported by the trend between argon deficiency
and \hi\ column density (Fig. \ref{arsnh}),
which indicates that local absorption does affect the level of ionisation.  
A possible example of local attenuation is provided
by the two DLAs\footnote{The DLAs at $z_\mathrm{abs}=2.402$ towards Q\,0027-1836
and at $z_\mathrm{abs}=2.309$ towards Q\,0100-1300, with
log N(\hi) $=21.75$ and $21.37$\,dex, respectively.}
with small [Ar/S] deficiency and low redshift that are visible 
in the left panel of Fig. \ref{arsz}. 
The high \hi\ column density of these two absorbers may provide
a strong \hi\ self-shielding, attenuating argon ionisation 
in a redshift interval where the external field is expected to be relatively strong.
If we exclude from our sample all DLAs with log N(\hi) $> 21.3$,
the Pearson's correlation coefficient of [Ar/S] versus $z_\mathrm{abs}$
increases from $\rho_\mathrm{Pe}$=+0.21 ($p_\mathrm{Pe}=42\%$)
to $\rho_\mathrm{Pe}$=+0.34 ($p_\mathrm{Pe}=26\%$). 
These considerations suggest that a weak trend with redshift 
may be present, but it is probably masked by local \hi\ self-shielding.

\begin{figure}
  \centering
\includegraphics[width=\columnwidth,clip=]{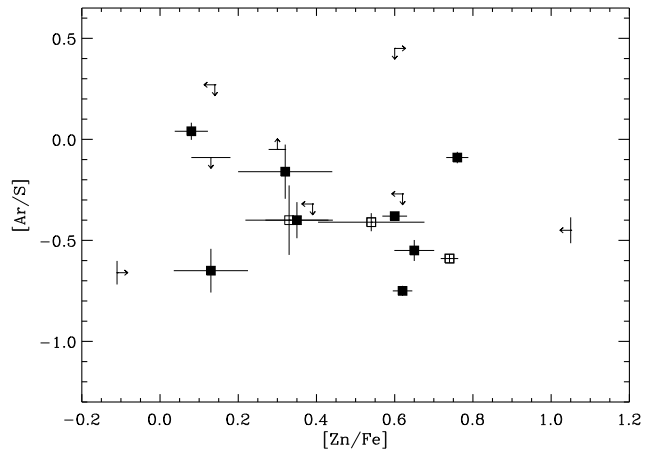}
     \caption{The [Zn/Fe] ratios against [Ar/S] abundance ratios in DLA systems. The open squares represent proximate DLAs.}
        \label{znfe}
 \end{figure}

\subsubsection{Probing the end of cosmic \heii\ reionisation}
In the upper range of the DLA redshift interval
the cosmic reionisation of \heii\ is expected to be completed \citep{furlanetto08,mcquinn09}.
When cosmic \heiii\ regions become completely overlapped, the photons with $h\nu > 54.4$\,eV
should travel undisturbed through the IGM, boosting the efficiency of argon ionisation processes.
The measurements versus redshift shown in Fig. \ref{arsz} do not show
a sudden rise of ionisation level that could be associated with the completion of \heii\ reionisation. 
Revealing this effect is quite difficult given the low statistics of DLA systems at high redshift. 
Our new data sample includes a total of 8 \ari\ measurements at $z_\mathrm{abs} \geq 3$
(5 more than in Paper\,I), but only one of these lies at $z_\mathrm{abs}>4$. 
Therefore, the redshift coverage of our sample is still marginal 
if one wants to follow the evolution of \heii\ reionisation, 
which apparently starts around $z \approx 6$ \citep{bolton12}
and is completed around $z \approx 3$
\citep{dallaglio08,syphers09,mcquinn09}.
At $z \leq 3$ our statistics is relatively large and suggests
that hard photons are able to ionise argon, with the exception
of the most self-shielded cases mentioned above.
This result is consistent with a scenario in which the IGM
becomes transparent to hard photons at some stage above $z=3$.
However, at high redshift, say $z>3.3$, we have only
two absorbers to probe earlier stages, and they give contrasting results. 
The DLA at $z_\mathrm{abs}=3.39$ towards Q\,0000-2620 \citep{molaro01} is characterized
by nearly solar argon abundances, whereas 
the DLA at $z_\mathrm{abs}=4.224$ towards J\,1443+2724 \citep{noterdaeme08}
shows evidence of argon deficiency. 
All together, our results 
are consistent with a scenario in which the \heii\ reionisation 
{\em is completed above} $z\simeq3$, but
a larger statistics of data at $z \geq 3.5$ is required to probe
the expected rise of IGM photons with $h\nu > 54.4$\,eV. 
 
\subsection{Testing the hypothesis of external ionisation}

The comparison of the argon abundances in DLA systems
with similar measurements in proximate DLA systems
and in the Galactic ISM provide ways of testing 
the hypothesis that the ionising field is external.

\subsubsection{Clues from proximate DLA systems}

If external ionisation dominates, we expect to find 
a strong argon deficiency in the proximate DLAs of our sample
due to their vicinity to a quasar source of external hard photons. 
%
In Fig. \ref{arsnh} we plot with different symbols 
the argon abundances versus N(\hi)
for the proximate DLAs (open squares) and the classic DLAs of our sample (filled squares). 
One can see that the three proximate absorbers 
have a stronger argon deficiency compared to the classic DLA systems 
of similar N(\hi).
This suggests that, for a given level of \hi\ self-shielding, 
the resulting argon ionisation is stronger in the proximate DLAs
due to the vicinity of the quasar. 
In Fig. \ref{arsz} one can see that the same three proximate absorbers
also have a stronger argon deficiency 
at a given redshift, 
compared to classic DLA systems. 
This suggests that, for a given level of metagalactic background
at a specific redshift,
their ionisation is stronger due to excess radiation of the nearby quasar. 
The analysis of the upper limits of [Ar/$\alpha$],
rather than the measurements, is consistent with these results.
In fact, the \ari\ upper limit for the proximate DLA system at $z_\mathrm{abs}=2.8754$ towards J\,1131+6044
\citep{ellison10} yields the strongest underabundance 
of any other absorber of our sample, with [Ar/Si] $<-1.06$\,dex. 

The results of these tests are in line with the hypothesis that argon ionisation
is very sensitive to external radiation with a quasar-like spectrum. 
Clearly, it is necessary to extend this test to a larger sample
in order to obtain a robust conclusion.
 
 \begin{figure}
 \centering
\includegraphics[width=\columnwidth,clip=]{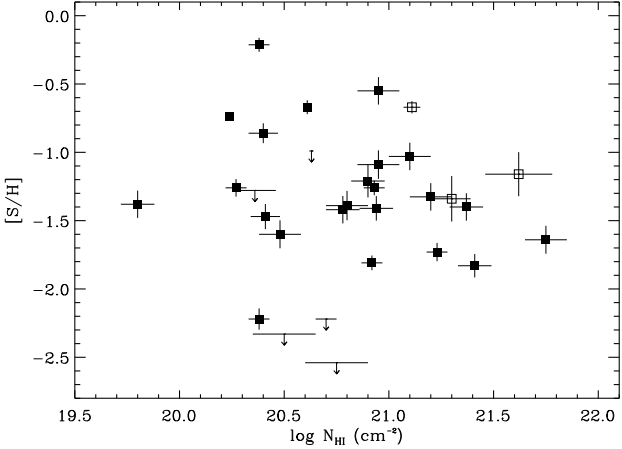}
   \caption{The [S/H] ratios versus \hi\ column density in DLA systems. The open squares represent proximate DLAs.}
       \label{shnh}
  \end{figure}

\subsubsection{Clues from Galactic interstellar studies}

A recent study of argon abundance in the Galactic ISM \citep{jenkins13}
indicates an average argon deficiency [\ari/\oi] $= -0.23 \pm 0.11$\,dex,
with the argon solar abundance\footnote{The ISM argon deficiency quoted by \citet{jenkins13} is
[\ari/\oi] $= -0.43 \pm 0.11$\,dex; this author adopts solar reference values
$\xi(\mathrm{O}) = 8.76$ and $\xi(\mathrm{Ar}) = 6.60$.} 
$\xi(\mathrm{Ar}) = 6.40$ adopted here \citep{asplund09}.
The mean [Ar/$\alpha$] that we measure in DLA systems
is $\simeq 0.2$\,dex lower than that measured in the ISM. 
The stronger underabundance in DLA systems is remarkable,
considering that the \hi\ column densities
of the Galactic lines of sight with \ari\ measurements
are significanty lower\footnote{
We estimate log N(\hi) $\leq 20$
for the Galactic lines of sight from the equivalent widths of the \oi\ lines published by \citet{jenkins13},
assuming solar abundances. 
The \ari\ absorption lines become saturated in a gas of solar composition
when log N(\hi) $\geq 20$.
Therefore it is practically impossible to compare \ari\ measurements
in ISM and DLA sightlines with same \hi\ column density. 
 }
than the \hi\ column densities in DLA systems. 
Therefore, \hi\ shielding of ionising radiation
should be less effective in the Galactic regions investigated by \citet{jenkins13}
than in the DLAs of our sample. 
As a result, 
for a given flux of ionising photons, argon should be more ionised in the ISM than in DLAs.
Since this is not the case, the flux of hard photons must be more intense in DLA systems 
than in the Galactic neutral gas. 
The comoving specific emissivity of IGM ionising photons
at the typical redshift of our sample ($2 \leq z \leq 4$)
is at least one order of magnitude higher than at present time 
\citep[see e.g. left panel of Fig. 15 in][]{haardt12}. 
The higher background at high redshift 
offers a natural explanation for the 
higher level of argon ionisation found in DLAs.
This explanation is consistent with the hypothesis that the ionising source of argon in DLAs is external.

\subsection{Internal sources of ionisation} \label{internalsources}
Finding sources of \ari\ ionisation internal to the DLA host galaxies faces the difficulty that
the spectrum of ionising photons
must include a significant hard component in order to explain the general level of argon deficiency (\S \ref{IonisationEffects}).
Stellar sources, characterized by relatively soft spectra, 
may contribute only to some extent to the observed level of argon ionisation. 
In principle, bubbles of hot interstellar gas ($T \geq 10^5$) created by supernova explosions could emit 
hard photons in the extreme UV/soft X. 
In fact, ionising sources of this type have been invoked by \citet{jenkins13} to explain
the argon deficiency observed in warm neutral regions of the Galactic ISM. 
In the case of the Galactic ISM, internal sources must be invoked
given the weakness of the quasar background at $z=0$. 
In our case this requirement is less compelling since the observed Ar deficiency can be reproduced assuming that the gas is ionised by the extragalactic radiation field at the redshift of the absorbers (\S\ref{IonisationEffects}).
%

\begin{figure}
  \centering
\includegraphics[width=\columnwidth,clip=]{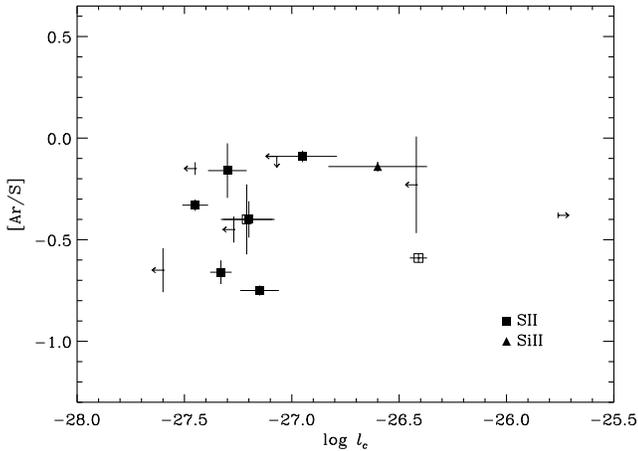}
     \caption{The [Ar/$\alpha$] ratios in DLA systems plotted versus cooling rate $\ell_c$
     (\S \ref{internalsources}). The squares and triangles correspond to the [Ar/S] and [Ar/Si] ratios, respectively. The open squares correspond to proximate DLAs.}
         \label{arscii}
  \end{figure}

Part of the DLA systems of our sample has a measurement of the \cii$^\ast$ column density,
which can be used to estimate the [\cii] 158 $\mu$m cooling rate of the gas \citep{wolfe08}, 
$\ell_c =$ N(\cii$^\ast$)/N(\hi) $A_{ul}h\nu_{ul}$, where N(\cii$^\ast$) is the column density of the
excited $^2P_{3/2}$ state in the $2s^2 2p$ term of C$^+$, $A_{ul}$ is the Einstein coefficient
for spontaneous photon decay to the ground $^2P_{1/2}$ state ($A_{ul} = 2.4 \times 10^{-6}$\,s$^{-1}$),
and $h\nu_{ul}$ is the energy of this transition ($h\nu_{ul}/k=92$\,K).
According to \citet{wolfe08}, DLAs show a bimodal distribution of $\ell_c$ values,
with two peaks at $\ell_c=10^{-27.4}$ and $\ell_c=10^{-26.6}$, representative of ``low-cool'' DLAs
and ``high-cool'' DLAs, respectively. 
In Fig. \ref{arscii} we plot the [Ar/S] and [Ar/Si] ratios versus $\ell_c$ for our sample. There are only 7 systems where \cii$^\ast$ has been measured and most of the \ari\ measurements (5 out 7) belong to the ``low'' $\ell_c$ group.  Even if the statistics at our disposal is quite small, one can see that in both DLA populations there are systems with both high and low Ar depletion: no trend is found between between Ar depletion and $\ell_c$. \citet{dutta14}, by means of CLOUDY models of metal-poor DLAs with measured \cii$^\ast$, and assuming a simple relation between SFR and CR ionisation rate, argued that in the ``low-cool" population the radiation field is similar to that of the Milky Way, while  in ``high cool''  DLAs  the radiation field is more than 4 times higher. The lack of a trend between Ar depletion and $\ell_c$ observed in our sample suggests that in situ SFR should not play a relevant role in determining the different degrees of Ar photoionisation.


\subsection{Interpretation of the trend with metallicity} \label{InterpretationMetallicity}
As the metallicity increases, cooling processes become more efficient, so that the gas should become less ionised and neutral argon more abundant. In line with this theoretical expectation, our CLOUDY model with HM extragalactic radiation (\S\ref{IonisationEffects}) predicts the [Ar/S] and [Ar/Si] ratios to increase by 0.04\,dex and 0.06\,dex, respectively, when the metallicity increases from 0.01 to 1.0 the solar reference level (case $z=2.5$, log N(\hi)$=20.75$\,dex). Even if the effect is almost undetectable, it is clear that the model does not predict the observed decrease of [Ar/$\alpha$] with metallicity (Fig. \ref{arsmet}, Tables \ref{correlations_full}-\ref{correlations}). This suggests that the idealized model of a gas layer directly exposed to the extragalactic background needs to be refined. The neutral gas layer intercepted by the line of sight will be, in general, surrounded by gas of the host galaxy that may absorb part of the extragalactic radiation. The trend with metallicity could be due to the different degree of transparency of the gas that surrounds the observed line of sight. The trend would be indirectly induced by the links that are known to exist between star formation efficiency and metallicity. The DLA galaxies of our sample span a relatively small interval of lookback time (top axis of Fig. \ref{arsz}), so that a higher metallicity implies a higher star formation efficiency. In turn, a higher star formation efficiency implies a stronger removal of neutral gas (astration) and a stronger production of ionised gas (supernova remnants). Since ionised gas is transparent to hard photons while neutral gas is not, DLAs in high-metallicity environments will be more directly exposed to the extragalactic radiation than DLAs in low metallicity environments. This scenario would yield a stronger \ari\ ionisation with increasing metallicity, in line with the observed trend. The trend may also be reinforced by local production of ionising photons from the hot gas heated by supernovae, as suggested by \citet{jenkins13} for the Galactic ISM.

\subsubsection{Bias effects}
 
An additional possibility to explain the trend 
with metallicity is the well-known
anti-correlation between metallicity and \hi\ column density
\citep{boisse98,vp05}.
As a result of this trend, DLA systems with high metallicity 
would experience more argon ionisation because they would have, on average,
low \hi\ self-shielding. 
In Fig. \ref{shnh} one can see that also in our sample 
the highest values of metallicity [S/H] are found at low \hi\ column density,
whereas systems of high \hi\ column density typically have low metallicities.
Clearly, this trend 
can contribute to the trend of argon deficiency with metallicity that we observe.
The existence of this bias makes less compelling, but does not rule out,
the interpretation based on the transparency of the DLA galaxies to ionising photons.

\section{Summary and conclusions}

We have used the EUADP quasar database \citep{zafar13,zafar13b} to search
for \ari\ lines at the absorption redshift of DLAs/sub-DLAs identified in the quasar spectra. 
We report 3 new measurements and 5 upper limits of \ari\ column density from the EUADP data.
We combine these data with the literature high-resolution measurements published to date. This combined sample comprises of 37 systems, i.e. the largest argon sample collected so far.
We have used this sample to investigate the relative abundance of argon in DLAs. 
%
%
%
The main results that we obtain can be summarized as follows.

\begin{itemize}
\item{Argon has a mean absolute abundance 
$\langle [\mathrm{Ar/H}] \rangle = -1.57 \pm 0.08$\,dex
(standard error of the mean),
lower than the mean metallicity of DLAs, [M/H] $\simeq -1.1$\,dex.
When compared with elements of similar nucleosynthetic origin, such as 
the $\alpha$-capture elements silicon and sulphur, we find a mean deficiency
$\langle [\mathrm{Ar/S}] \rangle = -0.39\pm0.05$\,dex
and $\langle [\mathrm{Ar/Si}] \rangle = -0.39\pm0.06$\,dex.
The scatter of these [Ar/$\alpha$] ratios is somewhat higher than the scatter
of other abundance ratios measured in DLAs, such as the [Zn/S] and [Si/Fe] ratios.}

\item{
Both [Ar/S] and [Ar/Si] ratios show evidence for
a positive correlation with \hi\ column density, N(\hi), and
absorption redshift, \zabs, and for a negative correlation with 
the dust-free metallicity, [S/H].
These trends are characterized by modest correlation coefficients,
the trend with redshift being particularly weak
(Tables \ref{correlations_full}-\ref{correlations}). 
}

\item{
Detailed analysis of the measured abundance ratios 
indicates that the argon deficiencies are due to ionisation processes,
rather than dust depletion or nucleosynthesis. 
The positive correlation between [Ar/$\alpha$] and log N(\hi)
is in line with the expectation that argon ionisation becomes
less important with increasing self-shielding by neutral hydrogen. 
The level of argon deficiency is broadly in agreement with the one predicted by simple models of DLA ionisation by extragalactic field at the redshift of the absorbers.
}

\item{
The behaviour of the Ar/$\alpha$ ratios with redshift and
N(\hi) can be interpreted as the combination of two factors: 
$i)$ a weakly evolving ionising background, 
with origin external to the DLA host galaxies, and $ii)$ an attenuation of the external radiation field according 
to \hi\ self-shielding of individual absorbers.
The combination of these two factors increases the scatter
of each one of the two correlations with N(\hi) and \zabs. 
}


\item{
The argon abundances of the few proximate DLAs present in our sample 
lie at the lower boundary of the abundances measured in classical DLAs
at the same \hi\ column density or same absorption redshift.
This result is consistent with the proposed scenario of argon ionisation by external quasar photons,
with enhanced ionisation of proximate DLAs
due to the vicinity of the background quasar. 
}

\item{
The average [Ar/$\alpha$] abundance in DLAs is $\simeq 0.2$\,dex lower than that found
in warm interstellar regions of the Milky Way \citet{jenkins13}.
The fact that the argon deficiency is stronger in DLAs
than in the Galactic ISM is consistent with the strong rise
of the metagalactic ionising background known to take place
from $z=0$ to $z\approx 3$, the typical redshift of our sample.

}

\item{
The negative correlation between [Ar/$\alpha$] ratios and metallicity that we find (Fig. \ref{arsmet}) cannot be explained with a simple model of gas layer directly exposed to the extragalactic background. We argue that DLAs in high-metallicity environments are likely to be more directly exposed to the extragalactic radiation than DLAs in low metallicity environments. Local production of ionising photons in hot gas might enhance argon ionisation in metal-rich environments.
}


\item{
Measurements of neutral argon in DLAs provide an independent test
of the evolution of ionising sources in the Universe.
Our measurements and upper limits of argon abundances suggest
that the \heii\ reionisation takes place 
at redshift higher than $z \sim 3$. The possibility of setting firmer constraints on the redshift evolution
of the \heii\ reionisation with our method requires a larger database of argon measurements at $z \geq 3.5$. }

\end{itemize}


\section*{Acknowledgements}
We would like to thank the ESO staff for making the UVES Advanced Data Products available to the community. We are thankful to the anonymous referee for his/her constructive comments. CP would like to thanks the BINGO! (`history of Baryons: INtergalactic medium/Galaxies cO-evolution') project by the Agence Nationale de la Recherche (ANR) under the allocation ANR-08-BLAN-0316-01. We are thankful P. Bonifacio for helpful discussions. G.V., M.C., and V.D. acknowledge support from the PRIN INAF ``The 1 Billion Year Old Universe: Probing Primordial Galaxies and the Intergalactic Medium at the Edge of Reionisation''. P. M. acknowledges the international team \#272 lead by C. M. Coppola ``EUROPA - Early Universe: Research on Plasma Astrochemistry'' at ISSI (International Space Science Institute) in Bern.

\bibliographystyle{aa}
\bibliography{argon.bib}{}

\bsp

\addtocounter{table}{-7}
\begin{landscape}
\begin{table}
\begin{center}                                 
\caption{Metal column densities and abundance ratios of DLA/sub-DLAs seen in \ari. The new estimates presented in this work are shown in bold-face. The proximate DLA cases are shown in italics.}           
\label{argon}    		
\setlength{\tabcolsep}{3pt}
\begin{tabular}{l c c c c c c c c c c c c c}
\hline\hline                       
QSO & \zem\ & \zabs\ & log N(\hi) & log N(\ari) & log N(\sii)  & log N(\siii) & log N(\feii) & Ref. & [Ar/S] & [Ar/Si] & [S/H] & [Zn/Fe] & log $f({\rm H_2})$\\
	 & & cm$^{-2}$ 	& cm$^{-2}$ & cm$^{-2}$ 	& cm$^{-2}$ & cm$^{-2}$ 	& \\ 
\hline
Q\,0000-2620 & 4.111 & 3.3901 & $21.41\pm0.08$ & $14.02\pm0.03$ & $14.70\pm0.03$ & $15.06\pm0.02$ & $14.87\pm0.03$ & 1 & $0.04\pm0.04$ & $0.07\pm0.04$ & $-1.83\pm0.09$ & $0.08\pm0.04$ & $\cdots$ \\
QXO\,0001 & $\cdots^{b}$ & 3.0000 & $20.70\pm0.05$ & $<13.38$ & $\cdots$ & $14.45\pm0.04$ & $<15.09$ & 2 & $\cdots$ & $<0.04$ & $\cdots$ & $\cdots$ & $\cdots$  \\
{\bf LBQS\,0010-0012} & 2.145 & 2.0250 & $20.95\pm0.10$ & $<14.17$ & $14.98\pm0.03$ & $15.31\pm0.05$ & $15.06\pm0.05$ & 3 & $<-0.09$ & $<-0.03$ & $-1.09\pm0.10$ & $0.13\pm0.05$ & $\cdots$ \\
B\,0027-1836 & 2.550 & 2.4020 & $21.75\pm0.10$ & $14.42\pm0.02$ & $15.23\pm$0.02 & $15.67\pm0.03$ & $14.97\pm0.02$ & 4 & $-0.09\pm0.03$ & $-0.14\pm0.04$ & $-1.64\pm0.10$ & $0.76\pm0.03$ & $-4.15\pm0.17$ \\
B\,0100+1300 & 2.686 & 2.3090 & $21.37\pm0.08$ & $14.21\pm0.12$ & $15.09\pm0.06$ & $>14.72$ & $15.09\pm0.01$ & 5 & $-0.16\pm0.13$ & $<0.60$ & $-1.40\pm0.10$ & $0.32\pm0.12$ & $\cdots$ \\
{\it J\,0140-0839} & 3.716 & 3.6940 & $20.75\pm0.15$ & $<12.82$ & $<13.33$ & $13.51\pm0.09$ & $<12.73$ & 6 & $<0.21$ & $<0.42$ & $<-2.54$ & $\cdots$ & $\cdots$ \\
{\it J\,0142+0023} & 3.370 & 3.3477 & $20.38\pm0.05$ & $<12.57$ & $13.28\pm0.06$ & $14.15\pm0.03$ & $\cdots$ & 6 & $<0.01$ & $<-0.47$ & $-2.22\pm0.08$ & $\cdots$ & $\cdots$ \\
Q\,0201+365 & 3.610 & 2.4630 & $20.38\pm0.05$ & $14.08\pm0.03$ & $15.29\pm0.01$ & $15.53\pm0.01$  & $15.01\pm0.01$ & 2 & $-0.49\pm0.03$ & $-0.35\pm0.03$ & $-0.21\pm0.05$ & $\cdots$ & $\cdots$ \\
B\,0336-017 & 3.197 & 3.0620 & $21.20\pm0.10$ & $>13.94$ & $14.99\pm0.01$ & $>15.14$ & $14.91\pm0.03$ & 2, 7 & $>-0.33$ & $>-0.09$ & $-1.33\pm0.10$ & $\cdots$ & $\cdots$ \\
B\,0347-383 & 3.222 & 3.0250 & $20.63\pm0.01$ & $13.99\pm0.02$ & $<14.76$ &  $15.24\pm0.01$ & $14.43\pm0.01$ & 7, 8, 9 & $>-0.05$ & $-0.14\pm0.02$ & $<-0.99$ & $0.30\pm0.02$ & $-5.80\pm0.07$ \\
B\,0450-1310B$^c$ & 2.250 & 2.0667 & $20.41\pm0.07^{a}$ & $<12.68^{a}$ & $14.06\pm0.06$ & $14.48\pm0.05$ & $14.13\pm0.02$ & 10 & $<-0.66$ & $<-0.69$ & $-1.47\pm0.09$ & $\cdots$ & $\cdots$ \\
{\it B\,0528-2505} & 2.756 & 2.8112 & $21.11\pm0.04$ & $14.25\pm0.01$ & $15.56\pm0.02$ & $16.01\pm0.03$ & $15.47\pm0.02$ & 9, 11, 12 & $-0.59\pm0.02$ & $-0.65\pm0.03$ & $-0.67\pm0.04$ & $0.74\pm0.02$ & $-2.59\pm0.16$ \\
{\bf B\,0642-5038} & 3.090 & 2.6590 & $20.65\pm0.06^{a}$ & $<14.61^{a}$ & $\cdots$ & $15.22\pm0.06$ & $14.91\pm0.03$ & 3, 9 & $\cdots$ & $<0.50$ & $\cdots$ & $\cdots$ & $-1.94\pm0.10$ \\
HS\,0741+4741$^c$ & 3.221 & 3.0170 & $20.48\pm0.10$ & $13.13\pm0.02$ & $14.00\pm0.02$ & $14.35\pm0.01$ & $14.05\pm0.01$ & 2, 7 & $-0.15\pm0.03$ & $-0.11\pm0.03$ & $-1.60\pm0.10$ &  $\cdots$ & $\cdots$  \\
B\,0841+129 & 2.495 & 2.3745 & $20.94\pm0.08^{a}$ & $13.53\pm0.08^a$ & $14.65\pm0.04$ & $15.16\pm0.03$ & $14.69\pm0.01$ & 10 & $-0.40\pm0.09$ & $-0.52\pm0.09$ & $-1.41\pm0.09$ & $0.35\pm0.08$ & $\cdots$ \\
B\,0841+129 & 2.495 & 2.4760 & $20.78\pm0.08$ & $13.11\pm0.09$ & $14.48\pm0.10$ & $14.99\pm0.03$ & $14.50\pm0.03$ & 10 & $-0.65\pm0.11$ & $-0.77\pm0.12$ & $-1.42\pm0.10$ & $0.13\pm0.09$ & $\cdots$  \\
Q\,0930+2858 & 3.425 & 3.2350 & $20.30\pm0.10$ & $<12.96$ & $\cdots$ & $13.89\pm0.02$ & $13.70\pm0.02$ & 2 & $\cdots$ & $<0.18$ & $\cdots$ & $\cdots$ & $\cdots$  \\
{\bf B\,1036-2257} & 3.103 & 2.7778 & $20.93\pm0.05$ & $<13.80$ & $14.79\pm0.02$ & $>14.97$ & $14.68\pm0.01$ & 3 & $<-0.27$ & $<-0.06$ & $-1.26\pm0.05$ & $<0.62$ & $\cdots$  \\
{\bf J\,1113-1533} & 3.370 & 3.2665 & $21.23\pm0.05^{a}$	& $13.55\pm0.06^a$ & $14.62\pm0.04$ & $>15.08$ & $14.65\pm0.03$ & 3 & $-0.35\pm0.07$ & $<-0.42$ & $-1.73\pm0.07$ & $\cdots$ & $\cdots$ \\
{\it J\,1131+6044} & 2.921 & 2.8754 & $20.50\pm0.15$ & $<12.52$ & $<13.29$ & $14.49\pm0.13$ & $>14.55$ & 6 & $<-0.05$ & $<-0.86$ & $<-2.33$ & $\cdots$ & $\cdots$ \\
J\,1211+0422 & 2.541 & 2.3766 & $20.80\pm0.10$ & $<13.10$ & $14.53\pm0.04$ & $14.91\pm0.04$ & $14.62\pm0.04$ & 13 & $<-0.71$ & $<-0.70$ & $-1.39\pm0.11$ & $\cdots$  & $\cdots$ \\
B\,1232+0815 & 2.570 & 2.3377 & $20.90\pm0.08$ & $13.86\pm0.22$ & $14.81\pm0.09$ & $15.06\pm0.05$ & $14.44\pm0.08$ & 9, 14 & $-0.23\pm0.09$ & $-0.09\pm0.23$ & $-1.21\pm0.12$ & $\cdots$ & $-1.03\pm0.18$ \\
{\it J\,1337+3152} & 3.174 & 3.1735 & $21.30\pm0.09^{a}$ & $13.96\pm0.10^{a}$ & $15.08\pm0.14$ & $15.44\pm0.08$ & $14.87\pm0.05$ & 15 & $-0.40\pm0.17$ & $-0.37\pm0.13$ & $-1.34\pm0.17$ & $0.33\pm0.11$ & $-7.11\pm0.17$ \\
J\,1340+1106$^c$ & 2.914 & 2.7958 & $20.92\pm0.05^{a}$ & $13.18\pm0.02^{a}$ & $14.23\pm0.02$ & $14.59\pm0.01$ & $14.23\pm0.01$ & 16 & $-0.33\pm0.03$ & $-0.30\pm0.02$ & $-1.81\pm0.05$ & $\cdots$ & $\cdots$ \\
BRI\,1346-03 & 3.992 & 3.7360 & $20.72\pm0.10$ & $<13.11$ & $\cdots$ & $13.95\pm0.01$ & $<14.13$ & 7 & $\cdots$ & $<0.28$ & $\cdots$ & $\cdots$ & $\cdots$ \\
{\bf J\,1356-1101} & 3.006 & 2.9669 & $20.80\pm0.10$ & $13.41\pm0.05$ & $\cdots$ & $14.95\pm0.07$ & $14.63\pm0.05$ & 3 & $\cdots$ & $-0.43\pm0.09$ & $\cdots$ & $<0.24$ & $\cdots$ \\
{\bf B\,1409+0930} & 2.838 & 2.6681 & $19.80\pm0.08$ & $<13.09$ & $13.54\pm0.06$ & $14.13\pm0.03$ & $14.02\pm0.03$ & 3 & $<0.27$ & $<0.07$ & $-1.38\pm0.10$ &  $<0.14$ & $\cdots$ \\
Q\,1425+6039 & 3.192 & 2.8268 & $20.30\pm0.04$ & $<13.43$ & $\cdots$ & $>14.83$ & $14.47 \pm0.01$ & 2 & $\cdots$ & $<-0.29$ & $\cdots$ & $0.65\pm0.01$ & $\cdots$  \\
J\,1443+2724 & 4.420 & 4.2240 & $20.95\pm0.10$ & $14.42\pm0.10$ & $15.52\pm0.01$ & $>15.43$ & $15.33\pm0.03$ & 7, 9, 17 & $-0.38\pm0.01$ & $>0.10$ & $-0.55\pm0.10$ & $0.60\pm0.03$ & $-2.36\pm0.18$  \\
{\it J\,1604+3951} & 3.130 & 3.1670 & $21.63\pm0.16^{a}$ & $14.45\pm0.03^{a}$ & $15.58\pm0.02$ & $15.96\pm0.02$ & $15.28\pm0.13$ & 6 & $-0.41\pm0.04$ & $-0.40\pm0.04$ & $-1.16\pm0.16$ & $0.54\pm0.14$ & $\cdots$ \\
GB\,1759+7539 & 3.050 & 2.6260 & $20.61\pm0.01^{a}$ & $13.68\pm0.03^{a}$ & $15.06\pm0.05$ & $15.40\pm0.02$ & $14.70\pm0.02$ & 18 & $-0.66\pm0.06$ & $-0.61\pm0.04$ & $-0.67\pm0.05$ & $>-0.11$ & $\cdots$  \\
HE\,2243-6031 & 3.010 & 2.3310 & $20.24\pm0.02^{a}$ & $13.35\pm0.05^{a}$ & $14.62\pm0.01$ & $14.93\pm0.02$ & $14.51\pm0.01$ & 19 & $-0.55\pm0.05$ & $-0.47\pm0.06$ & $-0.74\pm0.02$ &  $0.65\pm0.05$ & $\cdots$ \\
{\it J\,2321+1421} & 2.538 & 2.5731 & $20.70\pm0.05$ & $<13.33$ & $<13.60$ & $14.45\pm0.04$ & $14.18\pm0.03$ & 6 & $<0.45$ & $<-0.01$ & $<-2.22$ & $>0.60$ & $\cdots$  \\
{\bf B\,2332-094} & 3.330 & 3.0572 & $20.27\pm0.07^{a}$ & $12.96\pm0.05^a$ & $14.13\pm0.04$ & $14.64\pm0.03$ & $14.06\pm0.03$ & 3 & $-0.45\pm0.06$ & $-0.57\pm0.06$ & $-1.26\pm0.06$ & $<1.05$ & $\cdots$  \\
{\bf B\,2342+3417} & 3.010 & 2.9092 & $21.10\pm0.10$ & $<14.15$ & $15.19\pm0.01$ & $15.57\pm0.04$ & $15.02\pm0.06$ & 2, 3 & $<-0.32$ & $<-0.31$ & $-1.03\pm0.10$ & $<0.39$ & $\cdots$ \\
B\,2343+125 & 2.763 & 2.4310 & $20.40\pm0.07$ & $13.19\pm0.01$ & $14.66\pm0.02$ & $15.15\pm0.03$ & $14.52\pm0.02$ & 4, 9 & $-0.75\pm0.03$ & $-0.85\pm0.03$ & $-0.86\pm0.07$ & $0.62\pm0.03$ & $-6.41\pm0.16$  \\
Q\,2344+1228 & 2.515 & 2.5380 & $20.36\pm0.10$ & $<13.26$ & $<14.20$ & $14.18\pm0.01$ & $14.03\pm0.03$ & 7 & $<-0.22$ & $<0.19$ & $<-1.28$ &  $\cdots$ & $\cdots$  \\
\hline
\end{tabular}
\end{center}
{\bf References:} 1: \citet{molaro01}; 2: \citet{prochaska02}; 3: This work; 4: \citet{noterdaeme07}; 5: \citet{dessauges03}; 6: \citet{ellison10}; 7: \citet{prochaska01}; 8: \citet{levshakov02}; 9: \citet{noterdaeme08}; 10: \citet{dessauges06}; 11: \citet{centurion03}; 12: \citet{vladilo03}; 13: \citet{lehner08}; 14: \citet{balashev11}; 15 : \citet{srianand10}; 16: \citet{cooke11}, Cooke (2014, priv. comm.); 17: \citet{ledoux06}; 18: \citet{prochaska02b}; 19: \citet{lopez02}\\
$^{a}$ Sum of voigt profile components seen in \ari. The total N(\hi) is scaled down to match the fraction of \sii\ or \siii\ measured in the \ari\ voigt profile components. \\
$^{b}$ Emission redshift, $z_{em}$, of the quasar is not reported by \citet{prochaska02}. \\
$^c$ Cases with measurements of \aliii/\alii\ ratio: B\,0450-1310 (log \aliii/\alii\ $=-1.10$\,dex); HS\,0741+4741 (log \aliii/\alii\ $=-0.66$\,dex); J\,1340+1106 (log \aliii/\alii\ $=-1.09$\,dex).
\end{table}
\end{landscape}

\label{lastpage}
\end{document}